\documentclass[journal]{IEEEtran}
\usepackage[font=small,skip=1pt]{caption}
\IEEEoverridecommandlockouts
\RequirePackage{etex}  

\usepackage{textcomp}
\usepackage{gensymb}
\usepackage{latexsym}
\usepackage[T1]{fontenc}
\usepackage{multirow}
\usepackage{graphicx}
\usepackage{varioref}
\usepackage{bbding}
\usepackage{array}
\usepackage{enumitem}
\usepackage{soul,color}
\usepackage{longtable}
\usepackage{etex}
\usepackage{algorithm}
\usepackage{algpseudocode}
\usepackage{setspace}
\usepackage{amsmath}
\usepackage{amsfonts}
\usepackage{caption}

\usepackage{hyperref} 
\usepackage{cleveref} 

\usepackage{bigstrut}

\ifCLASSOPTIONcompsoc
  \usepackage[caption=false,font=normalsize,labelfont=sf,textfont=sf]{subfig}
\else
  \usepackage[caption=false,font=footnotesize]{subfig}
\fi

\usepackage{color}
\definecolor{light}{rgb}{0.5, 0.5, 0.5}

\medskip

\usepackage{epsfig,amssymb}

\usepackage{mathtools,amssymb}
\usepackage{lipsum} 

\usepackage{xcolor}
\usepackage[active]{srcltx} 
\usepackage{epstopdf} 

\usepackage{bm}

\usepackage{enumitem}
\DeclareRobustCommand*{\IEEEauthorrefmark}[1]{%
  \raisebox{0pt}[0pt][0pt]{\textsuperscript{\footnotesize\ensuremath{#1}}}}

\usepackage{cite}
\usepackage{graphicx}
\usepackage{textcomp}
\usepackage{xcolor}
\def\BibTeX{{\rm B\kern-.05em{\sc i\kern-.025em b}\kern-.08em
    T\kern-.1667em\lower.7ex\hbox{E}\kern-.125emX}}
\usepackage{lineno}
\usepackage[utf8]{inputenc}
\usepackage[english]{babel}
\usepackage{mathdots}

\graphicspath{{Pictures/}{../jpeg/}} 
\usepackage{float}

\begin{document}
\let\WriteBookmarks\relax
\def\floatpagepagefraction{1}
\def\textpagefraction{.001}

\title{Survey on Human-Vehicle Interactions and AI Collaboration for Optimal Decision-Making in Automated Driving}

\author{\IEEEauthorblockN{Abu Jafar Md Muzahid\IEEEauthorrefmark{1},
Xiaopeng Zhao\IEEEauthorrefmark{2},
and
Zhenbo Wang\IEEEauthorrefmark{3}}
\\
\vspace{1.5ex}
\IEEEauthorblockA{\IEEEauthorrefmark{1,2,3} Department of Mechanical, Aerospace, and Biomedical Engineering, The University of Tennessee, Knoxville, TN 37996, USA}\\
\thanks{ Corresponding author: Abu Jafar Md Muzahid (email: amuzahid@vols.utk.edu).}}

\markboth{Journal of \LaTeX\ Class Files,~Vol.~XX, No.~XX, XXXX~XXXX}{Shell \MakeLowercase{\textit{et al.}}: Bare Demo of IEEEtran.cls for IEEE Transactions on Magnetics Journals}

\maketitle
\begin{abstract} The capabilities of automated vehicles are advancing rapidly, yet achieving full autonomy remains a significant challenge, requiring ongoing human cognition in decision-making processes. Incorporating human cognition into control algorithms has become increasingly important, as researchers work to develop strategies that minimize conflicts between human drivers and AI systems. Despite notable progress, many challenges persist, underscoring the need for further innovation and refinement in this field. This review covers recent progress in human-vehicle interaction (HVI) and AI collaboration for vehicle control. First, we start by looking at how HVI has evolved, pointing out key developments and identifying persistent problems. Second, we discuss the existing techniques, including methods for integrating human intuition and cognition into decision-making processes and developing systems that can mimic human behavior to enable optimal driving strategies and achieve safer and more efficienttransportation. This review aims to contribute to the development of more effective and adaptive automated driving systems by enhancing human-AI collaboration.

\end{abstract}

\section{Introduction}
\label{sec:1}

Recent advancements in automated vehicle technology have dramatically transformed traffic dynamics, pushing the boundaries of modern transportation \cite{r3,r5}. From initial driver-assistance systems (DAS) to cutting-edge AI-driven models, these advancements have shown significant potential. However, achieving full autonomy remains a distant aim due to persistent challenges in safety, reliability, and decision-making that require ongoing human involvement \cite{r2}. Human cognition—characterized by adaptability, situational awareness, and intuitive decision-making—plays a crucial role in managing the complexities of real-world environments, which current AI models often struggle to replicate.

\begin{figure}[b]
    \centering
    \includegraphics[width=\linewidth]{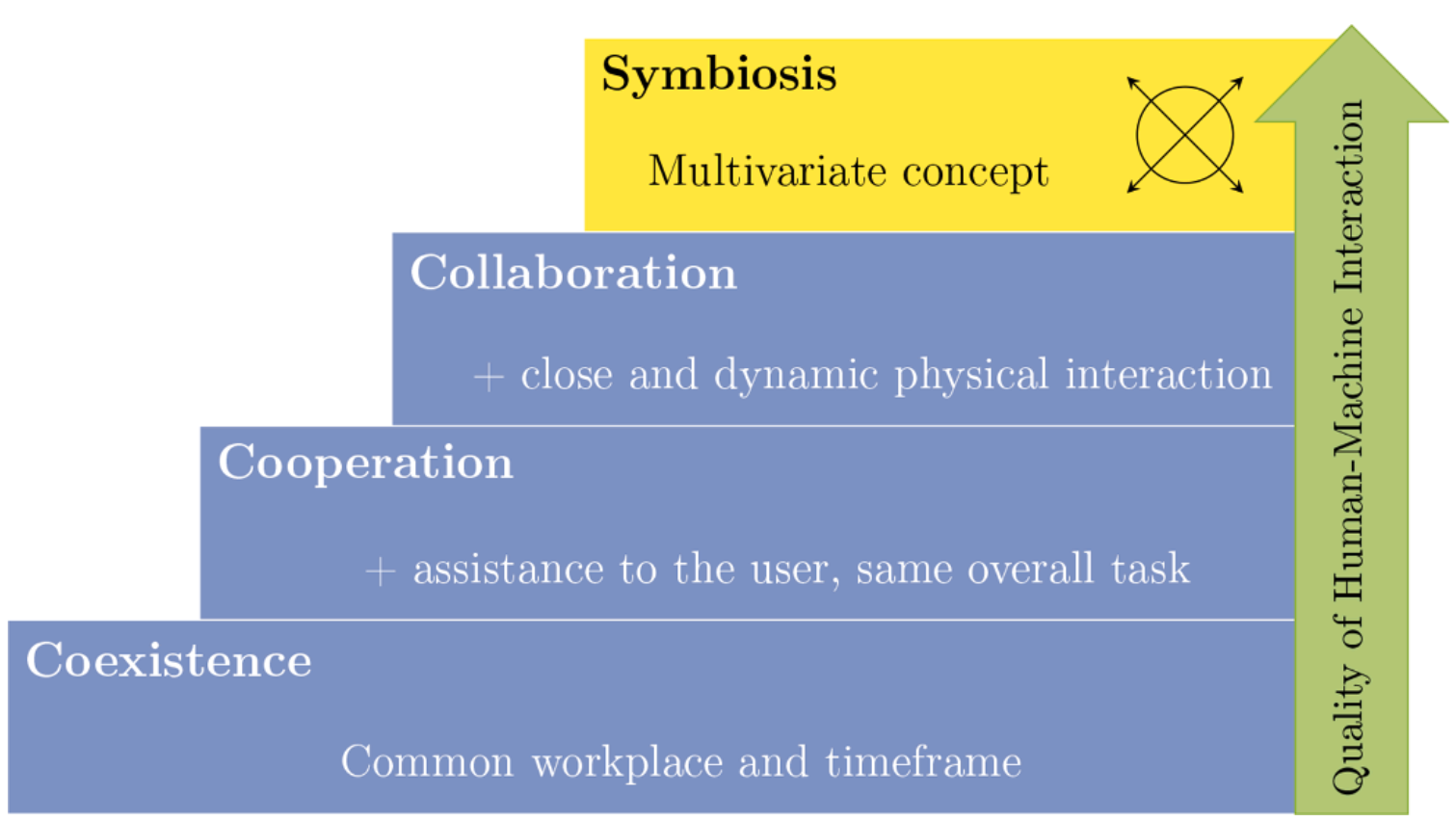}
    \caption{Levels of Human-Machine Interaction (HMI) from coexistence to symbiosis \cite{t77}.}
    \label{fig:symbiosis}
\end{figure}

To address these challenges, a symbiotic relationship between human cognition and automated systems is essential, as depicted in Figure~\ref{fig:symbiosis}. The figure illustrates the progressive levels of Human-Machine Interaction (HMI), starting with coexistence, where humans and machines share the same environment, advancing through cooperation and collaboration, and ultimately culminating in symbiosis. This multivariate concept emphasizes the integration of human strengths with machine efficiency to enhance decision-making quality, particularly in dynamic and safety-critical contexts like automated driving systems \cite{r7}. This manuscript critically examines the strengths and weaknesses of existing approaches to HVIs and AI integration, highlighting the unresolved gaps that must be addressed to realize the full potential of automated driving systems \cite{r3}.

In recent years, several review articles have significantly advanced the understanding of human-vehicle interactions (HVIs), particularly in decision-making and AI collaboration. For instance, \cite{t69} explored the future development of autonomous vehicles through a cognitive systems perspective, establishing foundational approaches for enhancing intelligent decision-making. Similarly, \cite{t70} examined hybrid intelligence systems, emphasizing their potential to improve decision-making in dynamic and nonlinear driving scenarios. The role of cognitive frameworks in enhancing system safety and efficiency was addressed in \cite{t71, t72}, which highlighted the integration of human-machine cognitive layers in autonomous driving. Furthermore, \cite{t73, t72} focused on the intersection of human and artificial intelligence, underscoring the importance of cognitive decision-making for achieving adaptive control in complex driving conditions. Collectively, these studies \cite{r3, r8, r123} provide a strong foundation for understanding the challenges and opportunities in modeling driver-vehicle interactions within the context of automation.

However, despite these valuable contributions, a critical gap remains in the literature—a focused analysis of traditional decision-making models tailored to address the unique challenges of Connected and Autonomous Vehicle (CAV) control systems. Existing reviews often fail to systematically integrate HVIs with cognitive modeling frameworks, leaving key aspects such as uncertainty modeling, operational constraints, and the interplay of HAC underexplored. This review addresses this gap by offering a comprehensive synthesis of cognitive decision-making models for CAV control strategies, focusing on their applicability, advantages, and limitations.

Through this investigation, we aim to bridge existing gaps by organizing and critically analyzing state-of-the-art techniques. Our work not only consolidates the current understanding of HVIs but also introduces a conceptual framework for decision-making layers and integration mechanisms. To further advance the field, we propose a generic algorithm structure for interpreting and optimizing HAC in autonomous driving. This review aspires to serve as a foundation for future innovations in cognitive decision-making and control strategies in automated systems.

This manuscript makes the following key contributions to the field of human-vehicle interaction (HVIs) and AI collaboration in automated driving:

\begin{enumerate}

    \item This manuscript critically examines the context of human-vehicle interactions (HVIs) and cognitive modeling, emphasizing the pivotal role of humans-in-the-loop in achieving effective collaboration. It explores advancements in cognitive modeling, particularly in understanding and simulating human behavior, and the development of innovative techniques that enhance adaptability and decision-making capabilities. Additionally, it provides a detailed analysis of traditional decision-making methods and cognitive models, critiquing their effectiveness and limitations in the context of automated driving.

    \item This review identifies and critically discusses the major challenges in cognitive modeling for HVIs and HAC. It highlights key issues related to understanding human behavior and intentions, addressing limitations in existing control strategies, and overcoming barriers to effective interaction. The manuscript provides a clear critique of these shortcomings and identifies specific areas where future research and development should focus.

    \item The manuscript proposes a novel taxonomy for HVIs and AI collaboration, categorizing levels of automation, decision-making layers, and integration mechanisms. It also introduces a conceptual framework to systematize decision-making and integration processes. To advance the field, it presents a generic algorithm structure designed to interpret and optimize HAC, offering practical insights for autonomous driving systems.

    \item Based on the critical findings, the manuscript provides targeted recommendations for future research and technological advancements. It outlines essential steps for refining cognitive models, improving human-machine communication, and enhancing AI capabilities, aiming to facilitate more effective human-vehicle collaboration and progress toward achieving full automation.

\end{enumerate}

The section \ref{sec:2} reviews the context of HVIs, emphasizing advances in cognitive modeling and their applications in modern systems. Section \ref{sec:3} explores key challenges in human-machine collaboration, including trust, communication, adaptability, and ethical considerations. Section \ref{sec:4} presents a taxonomy and conceptual framework for HVIs, focusing on decision-making layers and integration mechanisms. Finally, the conclusion summarizes the findings and outlines a roadmap for future research to advance HAC in automated driving systems.

\begin{figure}[ht]
    \centering
    \includegraphics[width=\linewidth]{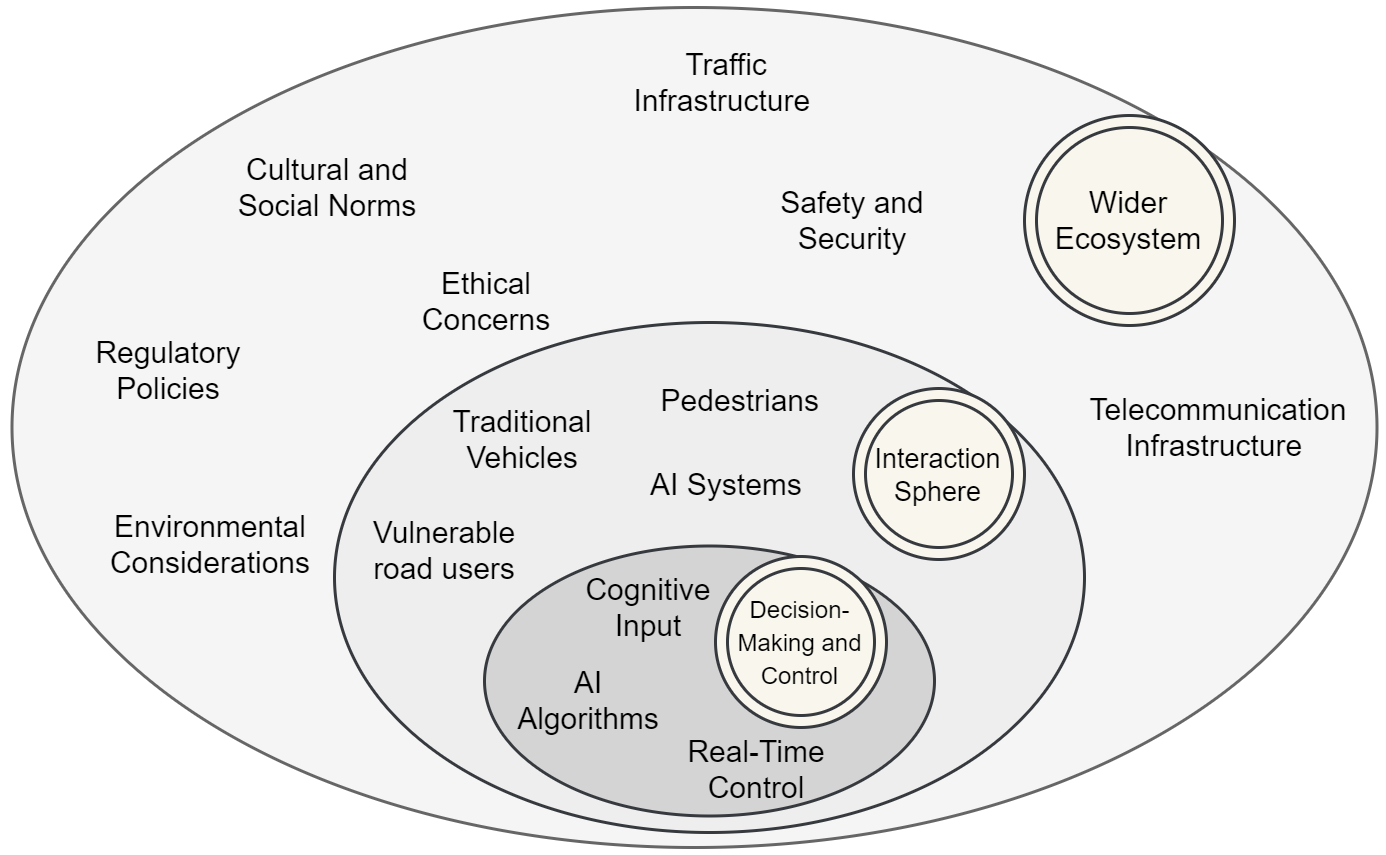}
    \caption{The Human-Vehicle Interaction Ecosystem, showing the interplay between cognitive inputs, AI systems, and external factors \cite{t73}.}
    \label{fig:ecosystem}
\end{figure}

\section{HVIs context and Cognitive Modeling}
\label{sec:2}
\subsection{The Role of Humans-in-the-Loop}
HVIs explores the intricate relationships between humans and automated vehicles (AVs), including drivers, cyclists, pedestrians, and other road users. As automation technology becomes increasingly integrated into vehicles, the complexity of these interactions grows, necessitating systems that account for human behavior, preferences, and expectations. H–AVI emphasizes the importance of seamless collaboration in diverse real-world environments, highlighting the critical need for humans to remain actively engaged in the system loop to ensure safety and adaptability \cite{s20}. Figure~\ref{fig:ecosystem} illustrates the multi-layered human-vehicle interaction ecosystem, where cognitive input, AI algorithms, and decision-making processes are influenced by broader contextual factors such as traffic infrastructure, environmental considerations, and regulatory policies. This highlights the interconnectedness of humans, AI systems, and the larger ecosystem.
\begin{figure}[!t]
    \centering
    \includegraphics[width=.8\linewidth]{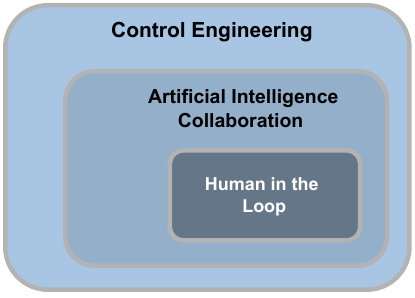}
    \caption{A layered framework of humans-in-the-loop, AI collaboration, and control engineering in HVIs \cite{t70}.}
    \label{fig:loop}
\end{figure}
Despite the rapid advancements in artificial intelligence (AI), human involvement is essential for addressing context-specific challenges and uncertainties. The concept of hybrid intelligence, which merges human adaptability with AI precision, illustrates the benefits of this collaboration. By fostering mutual learning and interaction, these systems can adapt to changing environmental dynamics and tackle complex challenges more effectively than either humans or AI could achieve independently \cite{s20}. This integration not only enhances the performance of automated systems but also ensures that they remain responsive to human needs and behavior. As shown in Figure~\ref{fig:loop}, humans-in-the-loop play a central role in the control architecture, with AI collaboration and control engineering forming layers that support and enhance human input.

Real-world applications of humans-in-the-loop approaches further demonstrate their value. For example, shared control systems enhance safety by combining human decision-making with the precision of automation \cite{s21}. These systems allow human operators to leverage their situational awareness while benefiting from the reliability of automated functions. Additionally, adaptive systems that monitor human workload can help mitigate risks associated with complacency and skill degradation in highly automated environments \cite{s22}. Cognitive modeling further enhances HVIs by simulating human behavior, enabling automated systems to predict and respond to human actions, intentions, and reactions in various driving contexts \cite{s23}. These models not only improve trust and adaptability but also create opportunities for designing systems that are more intuitive and responsive to user needs. Together, these advancements highlight the essential role of humans in the loop, not as mere participants but as integral components of systems designed to navigate the complexities of automated driving.

\begin{table*}[ht]
\centering
\caption{Comparison of Decision-Making Methods and Cognitive Modeling Approaches}
\begin{tabular}{|p{1.7cm}|p{7.6cm}|p{7.6cm}|}
\hline
\textbf{Aspect} & \textbf{Decision-Making Methods} & \textbf{Cognitive Modeling Approaches} \\ \hline
\textbf{Scope} & Focuses on algorithmic strategies for solving problems under uncertainty, often in structured environments. & Simulates human cognitive processes to enable human-like interaction, emphasizing adaptability to real-world complexities. \\ \hline
\textbf{Primary Goal} & Optimizing decisions based on mathematical frameworks (e.g., probabilities, rewards), prioritizing efficiency and effectiveness. & Enhancing HAC by modeling behavior, emotions, and intentions for intuitive interactions \cite{s97}. \\ \hline
\textbf{Techniques} & Includes finite state machines, fuzzy reasoning, Bayesian networks, reinforcement learning, imitation learning, and game theory. & Encompasses ACT-R, Soar, behavioral Cloning, Cognitive Digital Twins, Hybrid Models, and Neural-Symbolic Systems \cite{s98}. \\ \hline
\textbf{Adaptability} & Strong adaptability to stochastic and dynamic environments, allowing for real-time decision-making. & Focuses on human behavior and cognitive states in real-world interactions, facilitating personalized and adaptive responses. \\ \hline
\textbf{Learning Paradigm} & Utilizes explicit rules or learns optimal policies through interactions with environments (e.g., Reinforcement Learning, Markov Decision Processes). & Combines human cognitive architectures with learning algorithms for behavior prediction and context-specific adaptability \cite{s100}. \\ \hline
\textbf{Application Context} & Focused on autonomous driving tasks, such as trajectory optimization, collision avoidance, and path planning under uncertainty. & Enhances human-automated vehicle collaboration, including driver readiness monitoring, predictive handovers, and trust-building mechanisms. \\ \hline
\textbf{Human Interaction} & Limited to indirect contributions, such as algorithm design and parameter tuning, with minimal consideration for human factors. & Directly models human behavior to enable intuitive and personalized HAC, fostering trust and seamless interaction. \\ \hline
\textbf{Strengths} & - Effective in optimizing decisions under constraints. \newline - Handles randomness and multi-agent scenarios effectively. & - Mimics complex human behaviors and cognitive processes. \newline - Facilitates trust and communication in human-machine systems, improving user experience. \\ \hline
\textbf{Limitations} & - Lacks consideration for human preferences, emotions, and social dynamics. \newline - Computationally expensive for real-time applications. & - Requires extensive training data for accurate modeling. \newline - May oversimplify cognitive processes, leading to potential inaccuracies in behavior prediction. \\ \hline
\textbf{Evaluation Metrics} & Adaptability to uncertainty, consideration of randomness, learning efficiency, and decision quality. & Accuracy in behavior prediction, ability to foster trust, real-world validation of interactions, and user satisfaction. \\ \hline
\textbf{Examples} & Finite state machines, reinforcement learning, Bayesian networks, imitation learning, and game-theoretic models. & ACT-R, Soar, Hybrid Models, Neural-Symbolic Systems, and Cognitive Digital Twins. \\ \hline
\end{tabular}
\label{tab:decision_vs_cognitive}
\end{table*}

\subsection{Role and Advances in Cognitive Modeling}

\subsubsection{Understanding and Simulating Human behavior}

Cognitive modeling is integral to enabling automated systems to predict and adapt to human actions, intentions, and reactions in driving scenarios. These models, rooted in cognitive architectures such as ACT-R and Soar, replicate human decision-making processes and are particularly effective in managing unpredictable driving conditions \cite{f1}. The integration of machine learning further enhances these models, allowing them to adapt to real-time data and improve predictive capabilities. This synergy between cognitive modeling and machine learning has been pivotal in advancing the responsiveness of automated systems in HVIs \cite{f10}.

A comparative analysis, as outlined in Table \ref{tab:decision_vs_cognitive}, illustrates the distinct advantages of cognitive modeling over traditional decision-making methods. Unlike algorithmic approaches that primarily optimize decisions within structured mathematical frameworks, cognitive modeling incorporates human behavioral and emotional states, enabling adaptive responses tailored to real-world complexities \cite{f11}. These distinctions underscore the versatility of cognitive modeling in fostering HAC, especially in dynamic environments where traditional models may falter \cite{f12}. For example, cognitive models can account for factors such as driver fatigue and distraction, which are often overlooked in conventional algorithms \cite{f13, f14}.

Key applications of cognitive modeling include:
\begin{itemize}
    \item \textbf{Predictive Handovers}: By assessing driver readiness and anticipating transitions, cognitive models ensure smooth handovers between human and automated control, minimizing latency and errors in critical moments \cite{f2, f15}.
    \item \textbf{Real-Time behavior Prediction}: These models play a vital role in enhancing safety by dynamically predicting risky situations and adapting responses to mitigate potential hazards \cite{f2, f16}.
    \item \textbf{Human-AI Collaborative Decision-Making}: Cognitive models facilitate shared decision-making, combining human intuition with machine precision to navigate complex scenarios, thereby increasing system reliability and user trust \cite{f3, f17}.
\end{itemize}

\subsubsection{Advances in Modeling Techniques}

The continual evolution of cognitive modeling methodologies reflects a growing emphasis on precision, adaptability, and user-centric design. Several recent advancements stand out:

\begin{itemize}
    \item \textbf{Hybrid Models}: The integration of rule-based systems with machine learning has resulted in hybrid models capable of balancing reliability in structured scenarios with adaptability to unstructured and dynamic behaviors. This dual capability allows hybrid models to address a broader range of challenges in HVIs \cite{f1, f18}.
    
    \item \textbf{Neuro-Cognitive Approaches}: Leveraging insights from neuroscience, modern cognitive models incorporate real-time physiological data such as EEG signals and eye-tracking metrics to monitor driver attention, fatigue, and cognitive load. This integration enhances situational awareness and enables more adaptive system behavior, particularly in high-stakes scenarios \cite{f4, f19}.
    
    \item \textbf{behavioral Cloning}: By mimicking human driving styles, behavioral cloning facilitates seamless human-AI interactions. These models allow automated systems to learn from human behavior, adapting to individual preferences and fostering intuitive collaboration \cite{f2, f20}. However, it is essential to note that while behavioral cloning can enhance performance, it may also lead to overfitting if not properly managed \cite{f21}.
\end{itemize}

Testing and validating these advancements require sophisticated simulation platforms capable of replicating the complexity of real-world scenarios.  Table \ref{tab:simulators} summarizes key simulators utilized in cognitive modeling research. Tools such as CARLA and TORCS provide open-source environments for simulating urban and dynamic driving conditions, enabling the evaluation of decision-making algorithms under realistic settings \cite{f22, f23}. Other simulators, like Gazebo and ADAPS, extend these capabilities to multi-agent systems and emergency scenarios, offering researchers the flexibility to test cognitive models under diverse conditions \cite{f22, f24}. These platforms are instrumental in bridging the gap between theoretical developments and practical applications, ensuring the robustness and scalability of cognitive models for HVIs.

\begin{table*}
\caption{Key Simulators for Cognitive Modeling in HVIs Research}
\centering
\small
\begin{tabular}{|p{.9cm}|p{1cm}|p{4.7cm}|p{4.7cm}|p{4.7cm}|}
\hline
\textbf{Ref.} & \textbf{Simulator} & \textbf{Key Features} & \textbf{Cognitive Applications} & \textbf{Remarks} \\
\hline
[1][2] & CARLA & Open-source, supports urban layouts, dynamic scenarios, and high fidelity. & Cognitive training for decision-making in urban and dynamic scenarios. & Highly adaptable for various automated driving systems; supports multiple scenarios. \\
\hline
[4][5] & TORCS & Racing simulator with open-source code; models autonomous driving algorithms. & Ideal for simulating fast-paced decision-making and reaction times in competitive environments. & Limited vehicle models but effective for algorithm verification. \\
\hline
[7][8] & CarSim & High-fidelity vehicle dynamics with environmental and aerodynamic input support. & Models cognitive processes linked to vehicle control and handling under variable conditions. & Expensive proprietary software, known for highly accurate dynamics models. \\
\hline
[10][11] & Gazebo & ROS-based multi-agent simulation, supports sensor integration and modular flexibility. & Simulates multi-agent interactions, enabling cognitive research on cooperation and coordination. & Versatile but requires setup for automated vehicle-specific tasks. \\
\hline
[13][14] & ADAPS & Framework for accident-principled testing with high-fidelity scenarios. & Enhances cognitive safety models for emergency decision-making. & Reduces learning iterations for novel safety strategies in highly specific use cases. \\
\hline
[16][17] & LGSVL Simulator & Real-world scenarios with LiDAR, radar integration for AV development. & Advances cognitive models for sensor fusion and situational awareness in AVs. & Tailored for highly accurate sensor emulation; suitable for urban and highway environments. \\
\hline
[19][20] & Apollo Simulation & Baidu’s AV simulator with integrated road scenarios and flexible sensor configurations. & Focuses on culturally influenced cognitive behaviors and localized decision-making. & Best suited for scenarios involving Asian traffic environments; limited beyond that region. \\
\hline
[22][23] & SUMO & Traffic simulation with support for large-scale road networks and vehicle dynamics. & Provides input for cognitive models studying traffic flow and congestion management. & High scalability but simplified dynamics compared to dedicated vehicle simulators. \\
\hline
[25][26] & PreScan & Sensor and ADAS testing simulator with detailed vehicle-to-environment interactions. & Validates cognitive models integrating sensor feedback into safety-critical systems. & Focuses on sensor behavior validation; strong integration with external hardware. \\
\hline
[28][29] & AirSim & Aerial vehicle simulation with AV extensions, supports ML algorithm testing and diverse environments. & Models complex cognitive interactions between aerial and ground-based vehicles. & Excellent for AI research but requires extensions for ground-based AV simulations. \\
\hline
\end{tabular}
\label{tab:simulators}
\end{table*}

\begin{table*}[ht]
\centering
\caption{Methodologies and Metrics in Human-AI Collaboration for HVIs}
\begin{tabular}{|p{2cm}|p{3.5cm}|p{2cm}|p{3cm}|p{5cm}|}
\hline
\textbf{Methodology} & \textbf{Human Interaction} & \textbf{Platform} & \textbf{Key Metrics} & \textbf{Remarks} \\ \hline
Shared Control (e.g., torque) & Steering conflicts, driver authority. & Simulators & Lateral deviation, steering effort, workload. & Balances human and AI control dynamically, improving safety and precision \cite{s107}. \\ \hline
Adaptive Lane-Keeping & Driver distraction, secondary task load. & Simulators & RMSE, TLC, mental workload. & Enhances safety under distracted driving conditions, minimizing lane deviation \cite{s108}. \\ \hline
Predictive Handover & Driver drowsiness, readiness monitoring. & AV Platforms & Reaction time, system trust metrics. & Ensures seamless transitions between human and AI control, reducing response latency ([109]). \\ \hline
Cognitive Feedback Systems & Emotion/state monitoring (EEG, sensors). & Connected Systems & Cognitive load, real-time feedback accuracy. & Reduces mental workload, improves adaptability, and prevents cognitive overload \cite{s110}. \\ \hline
Driver-Intent Prediction & Driver gaze, hand positioning, pedal usage. & Simulators, AV Platforms & Predictive accuracy, driver intent recognition. & Enhances proactive safety measures and reduces ambiguity in shared decision-making \cite{s111}. \\ \hline
Autonomous Emergency Braking & Driver intervention during emergencies. & On-Road Vehicles & Braking delay, driver intervention rate. & Ensures timely responses in critical situations, supporting human oversight. \\ \hline
Real-Time Personalization & Adaptive to driver stress and preferences. & Connected Systems & Stress levels, user satisfaction, adaptability. & Aligns AI responses with individual driver preferences for enhanced trust and comfort \cite{s113}. \\ \hline
Haptic Feedback Systems & Physical feedback for steering or braking. & Simulators & Torque error, response consistency. & Reinforces driver understanding and collaboration with automated systems \cite{s114}. \\ \hline
Multimodal Communication & Verbal, visual, and tactile feedback for drivers and road users. & AV Platforms & Communication clarity, trust metrics. & Improves AV-human interaction, especially for pedestrians and cyclists \cite{s115}. \\ \hline
Human-State Monitoring & Fatigue, emotion, and alertness tracking. & Connected Systems & Fatigue detection rate, driver state accuracy. & Prevents fatigue-related accidents and aligns automated responses to human readiness \cite{s116}. \\ \hline
\end{tabular}
\label{tab:human_ai_metrics}
\end{table*}

\begin{table*}[ht]
\caption{Key Datasets for Cognitive Modeling in HVIs Research}
\centering
\small
\begin{tabular}{|p{1cm}|p{1.5cm}|p{1.5cm}|p{4cm}|p{4cm}|p{4cm}|}
\hline
\textbf{Ref.} & \textbf{Dataset} & \textbf{Category} & \textbf{Key Features} & \textbf{Cognitive Applications} & \textbf{Remarks} \\
\hline
\cite{s156} & NGSIM & Driving Dataset & High-precision raw video, lane changes, and extensive coverage of traffic scenarios. & Models driver behavior under various traffic conditions; useful for cognitive load analysis. & Broad coverage but limited to specific road scenarios, primarily focused on freeway environments. \\
\hline
\cite{s157} & HighD & Driving Dataset & Accurate vehicle trajectories with high temporal resolution. & Supports cognitive models for trajectory prediction and risk assessment in high-speed scenarios. & Limited focus on urban environments; highly detailed for highway data, making it less applicable to city driving contexts. \\
\hline
\cite{s158} & INTERACTION & Driving Dataset & High-resolution semantic maps and data on crisis situations. & Facilitates cognitive models of human reactions under stress and complex decision-making scenarios. & Well-suited for studying unpredictable real-world behaviors, particularly in emergency situations. \\
\hline
\cite{s159} & Lyft Dataset & Traffic Dataset & Detailed logs of traffic participant behavior, including interactions and routes. & Enables cognitive modeling of multi-agent interactions in urban traffic environments. & Focused on shared urban driving patterns, but limited in geographic coverage. \\
\hline
\cite{s160} & HDD & Driving Dataset & Longitudinal recordings of driving scenes with driver state mapping. & Enhances cognitive models by linking driving events to emotional and attentional states. & Limited to specific regions; suitable for long-term behavior studies and understanding driver fatigue. \\
\hline
\cite{s161} & Argoverse & Driving Dataset & Rich semantic map information, including lane annotations and static obstacles. & Supports trajectory prediction in cognitive models for spatial awareness and navigation. & High usability for lane-keeping and urban scenarios, providing a comprehensive view of driving environments. \\
\hline
\cite{s162} & nuPlan & Driving Dataset & Benchmarks and simulated traffic environments for autonomous vehicles. & Trains cognitive models for adaptive planning and decision-making under dynamic conditions. & Highly adaptable for training neural networks in structured environments, facilitating robust model development. \\
\hline
\cite{s163} & CommonRoad & Driving Dataset & Focused on path planning for autonomous vehicles, providing diverse scenarios. & Useful for modeling cognitive processes in automated trajectory optimization. & Supports diverse test cases but lacks real-world unpredictability, limiting its application in dynamic environments. \\
\hline
\cite{s164} & BDD100K & Multi-modal Dataset & Multi-camera dataset capturing diverse weather and lighting conditions. & Advances cognitive models in perception, enabling robust generalization to varied scenarios. & One of the most diverse datasets, facilitating robust generalization across different driving conditions. \\
\hline
\cite{s165} & Waymo Open Dataset & Traffic Dataset & High-resolution sensor data (LiDAR, radar, cameras) for comprehensive scene understanding. & Provides rich data for cognitive systems integrating multiple sensory inputs for decision-making. & Extremely high-resolution data; well-suited for large-scale automated driving experiments, enhancing model accuracy and reliability. \\
\hline
\end{tabular}
\label{tab:key-datasets}
\end{table*}

\subsection{Relevance to Modern HVI Systems}

The integration of cognitive modeling into HVI)systems has become indispensable for advancing safety, user experience, and system reliability. By enabling automated systems to predict and adapt to user actions and intentions, cognitive modeling addresses the complexities of human behavior in dynamic driving scenarios. This adaptability is particularly critical in semi-automated driving, where drivers frequently engage in non-driving tasks, resulting in potential lapses in attention and situational awareness. Research consistently highlights that cognitive load significantly influences driver performance, underscoring the need for systems that can dynamically adjust to users' cognitive states \cite{n25, n26}.

A major breakthrough in HVI systems is the development of adaptive human-machine interfaces (HMIs) that respond to drivers' cognitive and physiological states in real time. Frameworks such as the Cognitive Human-Machine Interface (CHMI) exemplify this approach, leveraging neurophysiological data—like EEG and heart rate—to optimize system responses based on driver workload and attention levels. This bidirectional interaction not only enhances situational awareness but also fosters a more intuitive relationship between humans and automation, ultimately improving driving performance and trust  \cite{n27, n28}. For example, methodologies such as \textit{cognitive feedback systems} (see Table~\ref{tab:human_ai_metrics}) use emotion and state monitoring to reduce mental workload and cognitive overload, reinforcing the system’s adaptability and user-centric design \cite{s110}.

While adaptive HMIs aim to ensure real-time responsiveness, cognitive modeling also plays a pivotal role in fostering calibrated trust in automated systems. Trust calibration is essential because misaligned trust—whether over-reliance or under-reliance—can lead to significant safety risks \cite{n29}. Cognitive models mitigate these risks by optimizing the complexity and clarity of information presented to users, thereby ensuring transparency and preventing automation misuse  \cite{n28}. Techniques such as \textit{real-time personalization}, detailed in Table~\ref{tab:human_ai_metrics}, align system behaviors with individual drivers' preferences and stress levels, creating systems that are both trustworthy and adaptive to unique user contexts  \cite{s113}.

\begin{figure}[ht]
    \centering
    \includegraphics[width=\linewidth]{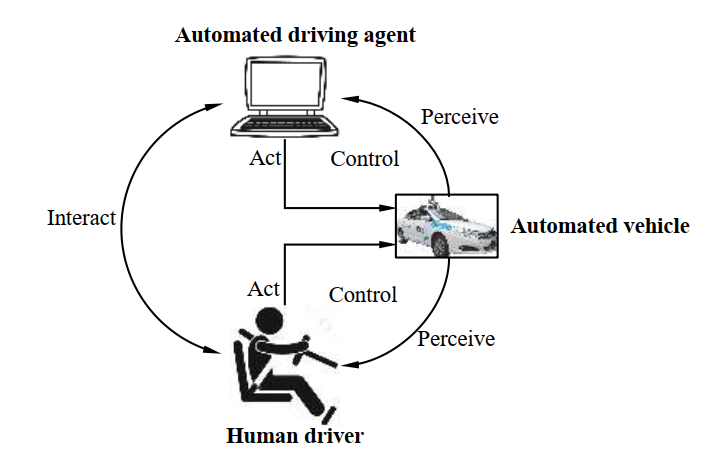}
    \caption{Cloning the human driving pattern \cite{t75}.}
    \label{fig-001}
\end{figure}

\subsubsection{Applications of Cognitive Modeling}

The diverse applications of cognitive modeling in HVI systems illustrate its transformative potential:
\begin{itemize}
    \item \textbf{Predictive Handovers}: By assessing driver readiness and predicting transitions between automated and manual control, cognitive models ensure smooth and error-free handovers during critical moments \cite{n30}.
    \item \textbf{Real-Time behavior Prediction}: These models dynamically identify risky situations, adapting system responses to mitigate potential hazards. This capability is crucial for improving driver confidence and maintaining engagement \cite{n31}.
    \item \textbf{Human-AI Collaborative Decision-Making}: Cognitive models facilitate shared decision-making, blending human intuition with machine precision. This collaboration enhances system reliability while fostering user trust \cite{n32}.
\end{itemize}

The practical deployment of these applications relies heavily on high-quality datasets that provide the necessary training and validation for cognitive models. Table~\ref{tab:key-datasets} summarizes some of the most critical datasets driving advancements in cognitive modeling. For instance:
\begin{itemize}
    \item \textbf{HighD Dataset}: Offers precise vehicle trajectory data for assessing high-speed scenarios, enabling cognitive models to anticipate risks in highway conditions \cite{s157}.
    \item \textbf{Waymo Open Dataset}: With its multimodal sensory inputs, this dataset supports the development of models capable of scene understanding across diverse environments \cite{s165}.
    \item \textbf{nuPlan Dataset}: Designed for simulated traffic environments, this dataset is instrumental in training models for decision-making and planning in dynamic conditions \cite{s162}.
\end{itemize}

These datasets enable cognitive models to generalize effectively, addressing challenges posed by diverse real-world conditions. However, gaps remain, particularly in representing underexplored driving contexts, such as varying geographic regions and cultural behaviors. Expanding dataset inclusivity will be crucial for achieving global adaptability in HVI systems.

\subsubsection{Advancements in Cognitive Modeling Methodologies}

The continual evolution of cognitive modeling methodologies underscores a growing emphasis on user-centric and adaptive design. Hybrid models, which integrate rule-based approaches with machine learning, strike a balance between reliability in structured scenarios and flexibility in dynamic environments. These models leverage structured frameworks for predictable interactions while using data-driven insights to adapt to emergent situations \cite{n33}. Neuro-cognitive approaches further refine this adaptability by integrating physiological data, such as eye-tracking and heart rate monitoring, to assess and respond to driver states in real time \cite{n34}.

However, implementing these advancements presents challenges that require careful consideration. While hybrid models reduce the limitations of purely rule-based systems, they also introduce computational complexity. Similarly, neuro-cognitive approaches demand the development of real-time monitoring systems that are both reliable and non-intrusive. The methodologies outlined in Table~\ref{tab:human_ai_metrics}, such as \textit{driver-intent prediction} and \textit{adaptive lane-keeping}, provide promising solutions by enhancing real-time responsiveness and safety \cite{s113}.

Datasets continue to be a cornerstone of progress, bridging the gap between theoretical advancements and practical deployment. For example, \textbf{INTERACTION Dataset} offers insights into multi-agent scenarios, while \textbf{BDD100K} captures diverse weather and lighting conditions, enabling cognitive models to handle edge cases effectively (Table~\ref{tab:key-datasets}) \cite{s158}.

Cognitive modeling is redefining the capabilities of modern HVI systems, making them safer, more adaptive, and increasingly human-centric. By integrating methodologies that dynamically adjust to human behavior (Table~\ref{tab:human_ai_metrics}) and leveraging datasets that capture diverse driving scenarios (Table~\ref{tab:key-datasets}), cognitive models are paving the way for more effective automation. However, achieving full autonomy will require addressing challenges related to ethical considerations, computational demands, and inclusivity in dataset design. As cognitive modeling continues to evolve, its integration into HVI systems promises to deliver not only enhanced safety and efficiency but also a deeper alignment with human needs, ensuring a future of seamless HAC.

\begin{figure*}[!t]
    \centering
    \includegraphics[width=\linewidth]{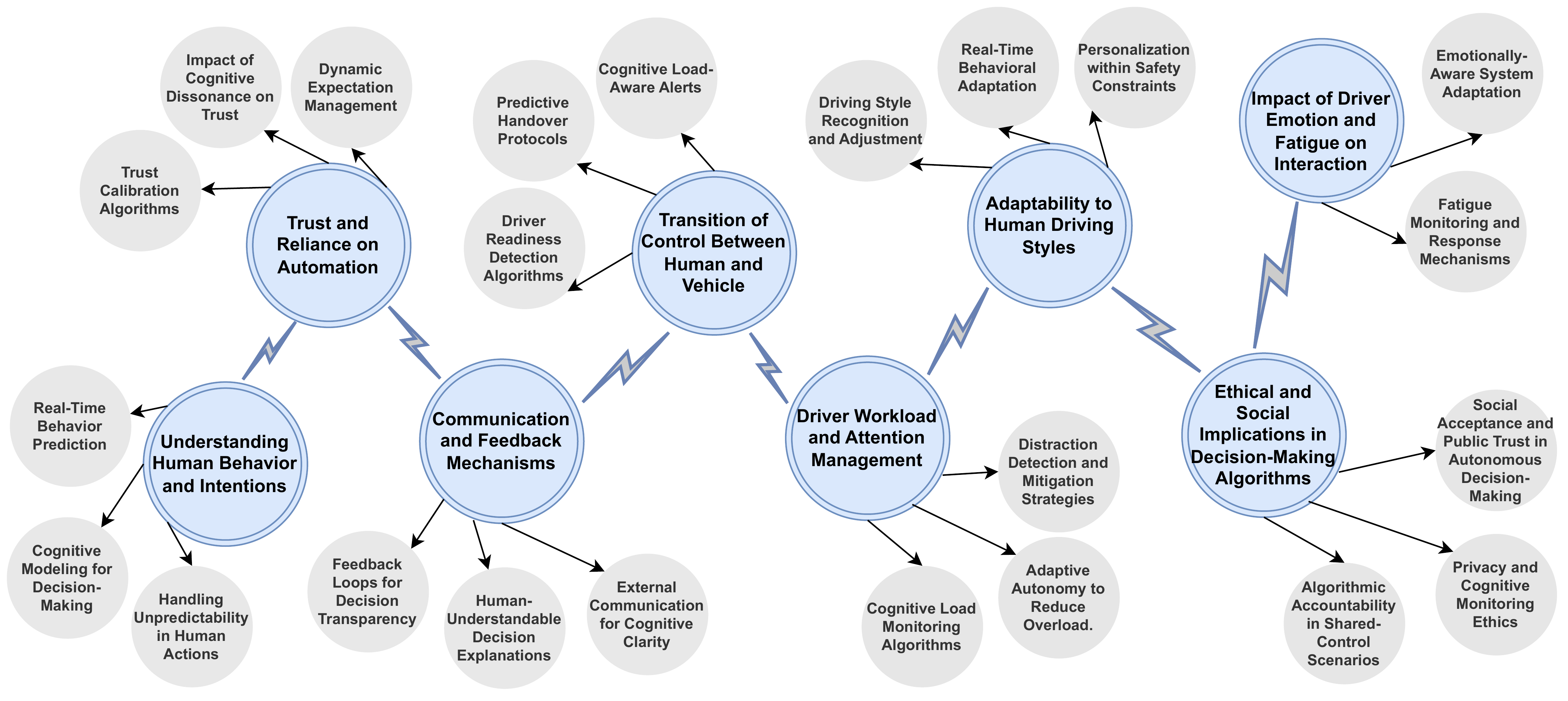}
    \caption{Key challenges in human-vehicle interactions (HVIs).}
    \label{fig:challenges}
\end{figure*}

\section{Challenges}
\label{sec:3}
The integration of human-vehicle interactions (HVIs) into automated driving systems presents several challenges that require careful consideration to ensure safety, efficiency, and trust. These challenges span various aspects, including understanding human behavior, maintaining trust in automation, managing communication and feedback mechanisms, and addressing ethical and social implications. Figure~\ref{fig:challenges} provides a visual overview of the key challenges, highlighting interconnected areas such as real-time behavior prediction, driver workload management, and adaptability to diverse driving styles. Additionally, issues like cognitive dissonance, transition control between humans and vehicles, and the impact of driver emotions on interaction are pivotal to enhancing collaboration between humans and machines. This section delves into these multifaceted challenges, offering insights into their complexity and emphasizing the need for targeted solutions to advance HVIs.
\subsection{Understanding Human behavior and Intentions}

\subsubsection{Real-Time behavior Prediction}
The ability to predict driver intentions in real-time is critical for enhancing the interaction between humans and automated vehicles. Developing algorithms that analyze cues such as gaze direction and hand movements can significantly improve the accuracy of predicting actions like lane changes or braking maneuvers. However, challenges remain in creating robust models that can consistently interpret these cues across diverse driving contexts and individual differences in behavior \cite{r121}. Research indicates that while progress has been made, the complexity of human behavior often leads to unpredictable outcomes that can compromise safety and efficiency in automated driving scenarios \cite{r122}.

\subsubsection{Cognitive Modeling for Decision-Making}
Integrating cognitive models that simulate human decision-making processes can enhance the responsiveness of HVI systems, especially in situations requiring rapid responses. These models must account for various factors influencing driver behavior, including stress, fatigue, and situational awareness \cite{r123}. Despite advancements in cognitive modeling, significant challenges persist in accurately replicating the nuances of human cognition, particularly under high-stress conditions where split-second decisions are critical \cite{r124}.

\begin{figure}[ht]
    \centering
    \includegraphics[width=\linewidth]{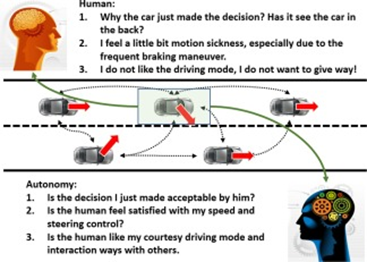}
    \caption{Illustrating potential decision-making conflicts between human drivers and autonomous systems \cite{t76}.
}
    \label{fig:conflicts}
\end{figure}

\subsubsection{Handling Unpredictability in Human Actions}
Human behavior is inherently unpredictable, posing a substantial challenge for automated systems designed to interact with drivers. Developing flexible decision algorithms that can adapt to a wide range of unpredictable human behaviors is essential for improving system robustness \cite{r125}. Figure~\ref{fig:conflicts} illustrates the potential conflicts between human drivers and autonomous systems, highlighting the mismatch in decision-making processes. For instance, humans may question the car's decisions or experience discomfort with its driving modes, while the autonomy system simultaneously evaluates whether its decisions align with human expectations and preferences. This dynamic underscores the need for adaptive systems capable of interpreting and responding to diverse human behaviors effectively. While machine learning techniques show promise in this area, the variability of human actions remains a significant hurdle, as algorithms must be trained on diverse datasets to generalize effectively across different driving scenarios \cite{r126}.

\subsection{Trust and Reliance on Automation}

\subsubsection{Trust Calibration Algorithms}
Adaptive algorithms that calibrate the level of assistance provided by automated systems based on the driver's trust level are crucial for fostering a healthy relationship between humans and automation. However, achieving this balance is challenging, as over-reliance on automation can lead to complacency, while mistrust can result in anxiety and disengagement \cite{r127}. Research indicates that while trust calibration mechanisms are being developed, there is still a lack of consensus on how to effectively implement these systems in real-world driving scenarios \cite{r128}.

\subsubsection{Impact of Cognitive Dissonance on Trust}
Cognitive dissonance occurs when there is a mismatch between driver expectations and the actions of the automated system, which can undermine trust. Addressing this dissonance is essential for maintaining user confidence in HVI systems \cite{r129}. Despite efforts to refine algorithms to minimize conflicts between system actions and driver expectations, significant challenges remain in ensuring that automated systems consistently meet user expectations, particularly in complex driving environments \cite{r130}.

\subsubsection{Dynamic Expectation Management}
Effective communication regarding the system's limitations and capabilities is vital for managing driver expectations. Algorithms that provide timely feedback about the automation's status can help drivers understand when they need to be more engaged or when they can rely on the system \cite{r131}. However, developing such dynamic expectation management systems is complex, as they must account for individual differences in driver cognition and experience with automation \cite{r132}.

\subsection{Communication and Feedback Mechanisms}

\subsubsection{Feedback Loops for Decision Transparency}
Implementing algorithms that enable real-time communication of the automated system's decisions to the driver can enhance cognitive alignment and trust. However, achieving transparency in decision-making processes remains a challenge, as drivers must be able to understand the rationale behind the system's actions \cite{r133}. Research suggests that while feedback mechanisms are being developed, many drivers still report feeling disconnected from the decision-making processes of automated systems, leading to reduced trust and engagement \cite{r134}.

\subsubsection{Human-Understandable Decision Explanations}
Simplifying algorithmic decisions into easily understandable feedback is essential for effective human-vehicle interaction. Despite progress in this area, many drivers still struggle to comprehend the system's behavior and intentions, particularly in complex driving situations \cite{r135}. This lack of clarity can undermine trust and lead to frustration, highlighting the need for ongoing research into effective communication strategies that bridge the gap between automated systems and human users \cite{r136}.

\subsubsection{External Communication for Cognitive Clarity}
In complex environments, external feedback signals for pedestrians and other road users can enhance safety by indicating vehicle intentions. However, the implementation of such communication mechanisms is fraught with challenges, as they must be designed to be intuitive and easily understood by all road users \cite{r137}. Research indicates that while external human-machine interfaces (eHMIs) are viewed as necessary for automated vehicles, there is still considerable uncertainty regarding their effectiveness in real-world scenarios \cite{r138}.

\subsection{Transition of Control Between Human and Vehicle}

\subsubsection{Driver Readiness Detection Algorithms}
Algorithms designed to assess driver alertness and readiness before transitioning control are essential for ensuring safe handovers between human and automated control. However, accurately gauging driver readiness remains a significant challenge, as it requires the integration of various biometric indicators and contextual factors \cite{r139}. Research shows that while progress has been made in this area, the variability of human responses to takeover requests complicates the development of reliable detection systems \cite{r140}.

\subsubsection{Predictive Handover Protocols}
Developing decision-making protocols that predict the optimal timing for control transfer based on real-time road and driver conditions is critical for enhancing safety. Despite advancements in predictive modeling, significant challenges remain in accurately assessing the myriad factors that influence the timing of control transfers \cite{r142}.

\subsubsection{Cognitive Load-Aware Alerts}
Alerts that consider the driver's cognitive load are vital for ensuring smooth handovers and reducing disorientation during control transfers \cite{r143}. Research suggests that while cognitive load-aware alerts can enhance driver readiness, there is still much to learn about how to effectively implement these systems in practice \cite{r144}.

\subsection{Driver Workload and Attention Management}

\subsubsection{Cognitive Load Monitoring Algorithms}
Systems that assess driver workload through biometric data can dynamically adjust the level of automation accordingly. However, accurately measuring cognitive load in real-time remains a significant challenge, as it requires sophisticated algorithms capable of interpreting complex physiological data \cite{r145}. Research indicates that while progress has been made in cognitive load monitoring, the variability of individual responses complicates the development of universally applicable systems \cite{r146}.

\subsubsection{Adaptive Autonomy to Reduce Overload/Underload}
Algorithms that dynamically modulate the system’s level of autonomy based on driver attention and engagement are essential for achieving an optimal cognitive balance \cite{r147}. Research shows that while adaptive autonomy can enhance driver engagement, significant challenges remain in ensuring that these systems operate effectively across diverse driving scenarios \cite{r148}.

\subsubsection{Distraction Detection and Mitigation Strategies}
Detecting driver distractions and intervening with prompts or adjustments in autonomy levels can help maintain a safe and engaged driving environment \cite{r149}. However, developing effective distraction detection systems is challenging, as they must accurately identify a wide range of distraction types and their impact on driving performance \cite{r150}.

\subsection{Adaptability to Human Driving Styles}

\subsubsection{Driving Style Recognition and Adjustment}
Algorithms that learn and adapt to the driver’s unique style can significantly influence the vehicle’s decision-making approach \cite{r151}. However, accurately recognizing and adapting to individual driving styles remains a challenge, as it requires sophisticated machine learning techniques capable of processing diverse behavioral data.

\subsubsection{Real-Time behavioral Adaptation}
Algorithms that adjust system responses based on the driver’s immediate actions offer a tailored driving experience while maintaining safety \cite{r153}. However, developing such real-time adaptation systems is complex, as they must continuously monitor driver behavior and adjust accordingly \cite{r154}.

\subsubsection{Personalization within Safety Constraints}
Balancing customization with safety limits is crucial in developing algorithms that adapt to driving style preferences without compromising decision-making integrity \cite{r155}. Research emphasizes the importance of maintaining safety as a priority while allowing for personalization in automated driving systems \cite{r156}.

\subsection{Ethical and Social Implications in Decision-Making Algorithms}

\subsubsection{Algorithmic Accountability in Shared-Control Scenarios}
Defining responsibility within algorithms during shared-control scenarios is essential for ensuring accountability in decision-making processes \cite{r157}. Research highlights the need for transparent accountability mechanisms to foster trust and acceptance of automated systems \cite{r158}.

\subsubsection{Privacy and Cognitive Monitoring Ethics}
Ensuring data privacy in cognitive monitoring systems that rely on biometric or behavioral data is critical for ethical considerations in HVI systems \cite{r159}. Research emphasizes the importance of ethical data practices in maintaining user trust and acceptance of automated technologies \cite{r160}.

\subsubsection{Social Acceptance and Public Trust in Autonomous Decision-Making}
Examining how cognitive transparency in algorithmic decisions influences public acceptance of automated systems is vital for fostering societal trust \cite{r161}. However, significant challenges remain in effectively communicating how decisions are made to enhance public confidence in automated driving technologies \cite{r162}.

\subsection{Impact of Driver Emotion and Fatigue on Interaction}

\subsubsection{Fatigue Monitoring and Response Mechanisms}
The impact of driver emotion and fatigue on interaction in automated vehicles is a critical area of study, particularly in the context of fatigue monitoring and response mechanisms. Systems designed to detect signs of driver fatigue are essential for ensuring safety in automated vehicles. The challenge of accurately monitoring fatigue levels in real-time is significant, as it necessitates the integration of various biometric indicators and contextual factors. For example, fatigue detection systems often rely on observable signs such as yawning, blink duration, and head movements, which can vary significantly among individuals. Recent advancements have been made in developing these systems; however, the variability of human responses complicates the creation of reliable detection systems. Studies indicate that biometric measures, such as heart rate variability and skin temperature, show promise in predicting fatigue, but their effectiveness can be influenced by external factors, including environmental conditions and individual differences \cite{r167, r168}. Thus, while progress is evident, achieving consistent and accurate fatigue monitoring in diverse driving scenarios remains a significant challenge.

\subsubsection{Emotionally-Aware System Adaptation}
Emotionally-aware system adaptation is another crucial aspect of enhancing driver-vehicle interactions. Algorithms that consider emotional states when adjusting vehicle behavior can significantly improve decision alignment with driver needs. However, developing such emotionally aware adaptations is complex, as they must accurately interpret a wide range of emotional cues and adjust system behavior accordingly. Research suggests that emotionally aware systems can enhance driver trust and satisfaction, but significant challenges remain in ensuring that these systems operate effectively in real-world scenarios \cite{r169}. For instance, detecting emotions through biometric signals, such as facial expressions or physiological responses, can be influenced by various factors, including the driver's current context and the presence of distractions \cite{r170}. This complexity underscores the need for ongoing research to refine these algorithms and ensure they can adapt to the dynamic nature of human emotions during driving.

While significant advancements have been made in the areas of fatigue monitoring and emotionally-aware adaptations, substantial challenges persist. The integration of biometric data and the development of algorithms that can accurately interpret this data in real-time are critical for enhancing the safety and effectiveness of human-vehicle interactions. Continued research is essential to address these challenges and improve the overall driving experience in automated vehicles.

\begin{figure*}[!t]
    \centering
    \includegraphics[width=.7\linewidth]{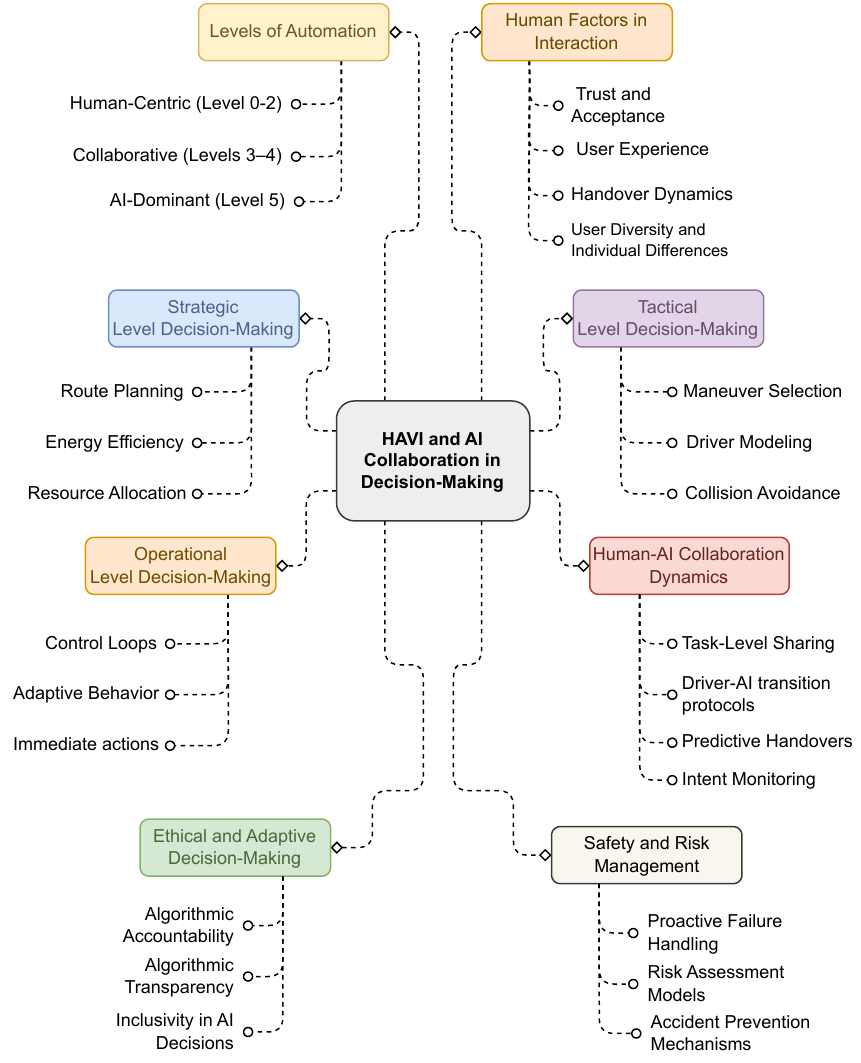}
    \caption{Taxonomy of HVI and AI Collaboration in Decision-Making.}
    \label{Taxonomy}
\end{figure*}

\section{Taxonomy and Conceptual Framework}
\label{sec:4}

\subsection{Taxonomy of HVI and AI Collaboration in Decision-Making}

The taxonomy of HVIs offers a comprehensive framework to categorize the interplay between humans and AI systems in automated driving contexts. As HVI technologies evolve, this taxonomy highlights key dimensions essential for understanding and enhancing HAC, enabling adaptive and effective decision-making. The hierarchical structure provided by the taxonomy presented \textit{Figure \ref{Taxonomy}} categorizes these dimensions into three interconnected themes: levels of automation, decision-making layers, and supporting components. Each theme plays a pivotal role in addressing the challenges and opportunities in HVI systems.

\subsubsection{Levels of Automation}

A foundational aspect of this taxonomy is its alignment with the progression of automation levels, as defined by the Society of Automotive Engineers (SAE). These levels, spanning from human-centric control to full AI dominance, demonstrate the shifting responsibility of decision-making between humans and machines \cite{t48}.

At the \textit{human-centric levels (0--2)}, the system primarily assists drivers through alerts and partial automation. However, this limited automation introduces challenges such as cognitive overload and decision fatigue, particularly during prolonged manual control. Monitoring cognitive workload becomes critical to ensure that drivers remain engaged and responsive, as highlighted by Payre et al. \cite{t49}. Transitioning to \textit{collaborative levels (3--4)}, shared control emerges as a defining feature, with systems and humans dynamically sharing decision-making responsibilities. Trust calibration and seamless handovers become crucial at this stage. Research by Kaber emphasizes the need for adaptive trust-building mechanisms to foster user confidence in these collaborative systems \cite{t50}. Finally, at the \textit{AI-dominant level (5)}, where the system assumes full control, the focus shifts to ethical considerations, inclusivity in decision-making, and maintaining accountability. Utriainen and Pöllänen underscore the importance of designing transparent decision-making processes that accommodate diverse societal norms and ethical frameworks \cite{t51}.

\subsubsection{Decision-Making Layers}

In parallel to the levels of automation, the taxonomy categorizes decision-making into three interdependent layers: strategic, tactical, and operational. Each layer represents a distinct domain of decision-making, contributing to the overall functionality of HVIs.

The \textit{strategic level} encompasses long-term decision-making tasks, such as route optimization and resource allocation. By integrating predictive models, systems at this layer can dynamically adapt to evolving conditions, such as traffic patterns or weather changes, while aligning with user preferences. Research demonstrates the value of cognitive adaptability in strategic planning, enabling systems to anticipate and mitigate potential challenges effectively \cite{t52, r12}. Transitioning to the \textit{tactical level}, decision-making focuses on real-time maneuvering, such as lane changes and speed adjustments. Tactical-level decisions rely on situational responsiveness, where evidence accumulation models enhance safety by reducing reaction times and improving decision precision \cite{t53}. Finally, the \textit{operational level} involves immediate actions, such as collision avoidance or emergency braking. Task-analytic models are instrumental in guiding these split-second decisions, ensuring reliability and robustness \cite{t54}.

\subsubsection{Supporting Components}

Complementing the levels of automation and decision-making layers are several supporting components that address human factors and system capabilities, ensuring seamless collaboration in HVIs.

\begin{itemize}
    \item \textbf{Human Factors in Interaction:} Trust, acceptance, and user experience are critical to successful HAC. Systems must account for individual differences and cultural diversity to enhance user satisfaction and engagement. Adaptive designs are essential for accommodating varying cognitive profiles and behaviors \cite{t52}.
    \item \textbf{Shared Control Systems:} These systems enable seamless collaboration by integrating intent monitoring, predictive handovers, and task-level sharing. Research highlights the role of predictive handover protocols in minimizing errors during transitions, improving overall system reliability \cite{t55}.
    \item \textbf{Ethical and Adaptive Decision-Making:} Addressing accountability and inclusivity in AI-driven decisions is paramount. Proactive risk management models that prioritize ethical considerations and fairness across user groups are essential \cite{t56}.
    \item \textbf{Safety and Risk Management:} Effective safety mechanisms rely on cognitive models for risk assessment and failure handling. These mechanisms anticipate hazards and adapt responses dynamically, mitigating risks in real time. Research has identified mode error and awareness as critical factors influencing safety in automated systems \cite{t57}.
\end{itemize}

\subsubsection{Transition and Integration}

This taxonomy provides a structured lens to understand the complex interactions in HVI systems. By categorizing key components into actionable layers, it bridges theoretical advancements with practical applications, guiding the development of adaptive, user-centric systems. The following subsection will extend this discussion by proposing a conceptual framework that operationalizes the taxonomy, translating its insights into actionable strategies for decision-making and collaboration in HVIs.

\begin{figure*}[ht]
    \centering
    \includegraphics[width=\linewidth]{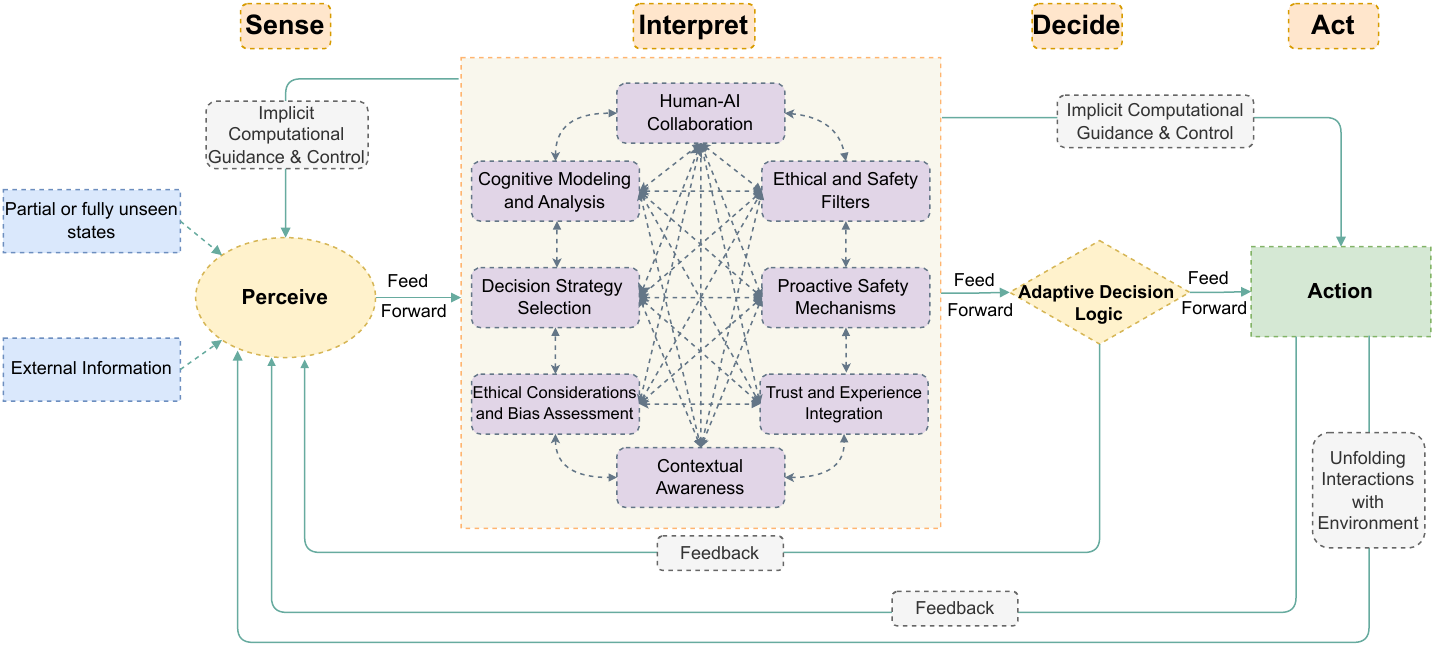}
    \caption{Conceptual Framework.}
    \label{Conceptual_Framework}
\end{figure*}
\subsection{Conceptual Framework for HVI and AI Collaboration}

Building on the taxonomy, this conceptual framework serves as an actionable guide to the design and implementation of human-vehicle interaction (HVI) systems. The framework illestrated by the \textit{Figure \ref{Conceptual_Framework}} integrates decision-making layers, cognitive modeling, and human-AI interaction into a unified system, emphasizing adaptability, transparency, and safety. By addressing key challenges, it bridges theoretical insights with practical applications.

\subsubsection{Sense-Interpret-Decide-Act (SIDA) Cycle}

At the heart of this framework lies the Sense-Interpret-Decide-Act (SIDA) cycle, a structured process that ensures a continuous and systematic flow of information for real-time decision-making. Each stage of the cycle contributes to the seamless integration of human cognition and AI capabilities.

\textbf{Sense:}

This stage encompasses the acquisition of data from diverse sources, including sensors, cameras, and environmental inputs. By translating raw information into actionable insights, the system gains awareness of both immediate and contextual factors, such as road conditions and user-specific behaviors. Recent studies highlight the importance of integrating multimodal sensor data to enhance situational awareness in automated vehicles \cite{t58}. For instance, Li et al. emphasize the significance of this integration in improving the accuracy of sensing systems.

\textbf{Interpret:}

Interpretation is achieved through cognitive modeling, where algorithms synthesize data to derive situational awareness. This step incorporates user preferences, emotional states, and historical data to provide a nuanced understanding of the driving context. Techniques such as Bayesian causal cognition and conceptual spaces enhance the ability to make sense of complex and ambiguous situations, improving the reliability of subsequent decision-making stages \cite{t59, r13}. Decision-making strategies integrate insights from the interpretation phase with ethical filters and contextual awareness. This stage balances human intuition with AI precision, particularly in scenarios requiring rapid yet fair and transparent responses. Advanced frameworks, such as autonomous cognitive entities, enable systems to prioritize safety while maintaining inclusivity in decision-making \cite{t60}.

\textbf{Act:}

The act stage involves executing decisions through adaptive control loops and safety mechanisms. Proactive safety measures, including predictive handovers and collision avoidance, ensure robust responses to dynamic changes in the driving environment. Research by Gold et al. highlights the importance of effective handover strategies to minimize risks during transitions between automated and manual control \cite{t59}. Additionally, the integration of feedback from previous actions allows the system to refine its responses continuously, enhancing overall safety and reliability \cite{t61, r14}.

\subsubsection{Key Components of the Framework}

To enhance the functionality of the SIDA cycle, the framework incorporates several critical components that address the unique challenges of HVIs:

\textbf{Trust and Experience Integration:}

User trust is integral to the success of HVI systems. By embedding user feedback and past interactions into decision algorithms, systems can dynamically adjust to individual needs, fostering trust and confidence. Multilayer cognitive networks, which model human behaviors and predict preferences, are particularly effective in reinforcing seamless collaboration between humans and AI \cite{t60}. Research by Körber et al. emphasizes that trust in automation is crucial for user acceptance and effective interaction with automated systems \cite{t62}.

\textbf{Ethical and Safety Filters:}

The framework incorporates filters to address biases, risks, and ethical considerations. By leveraging models of bounded rationality, systems balance ethical imperatives with real-time operational demands, ensuring fairness across diverse scenarios \cite{t63}. This is particularly important in automated driving, where ethical dilemmas can arise during decision-making processes \cite{t63}.

\textbf{Adaptive Decision Logic:}

Real-time adaptability is a cornerstone of the framework. Systems utilize high-dimensional conceptual spaces to refine decision-making, accommodating evolving environmental and user-specific factors. Research indicates that adaptive decision-making strategies can significantly enhance the responsiveness of HVI systems to changing conditions \cite{t64}.

\subsubsection{Feedback and Feedforward Mechanisms}

\begin{itemize}
    \item \textbf{Feedback Mechanisms:} These mechanisms capture real-time system outcomes, allowing decision strategies to evolve through iterative learning. Feedback is crucial for refining system performance and adapting to unforeseen challenges \cite{t61}.
    \item \textbf{Feedforward Mechanisms:} Anticipatory in nature, feedforward mechanisms enable the system to predict future states and preemptively adjust its strategies. These mechanisms leverage predictive analytics and causal representation learning to minimize potential risks \cite{t65}.
\end{itemize}

\subsubsection{Human-AI Collaboration}

Collaboration is embedded at every level of the framework, ensuring that AI capabilities complement human expertise. The framework emphasizes:

\begin{itemize}
    \item \textbf{Shared Control:} Human inputs are integrated into task-level decisions, fostering a partnership where human intuition and AI precision work in tandem. Research by Stoll et al. supports the concept of shared control as a means to enhance safety during transitions \cite{t66}.
    \item \textbf{Transparency and Trust:} Interpretable models enhance user understanding of system behavior, promoting engagement and acceptance. Studies show that transparency in AI decision-making processes is vital for building user trust.
    \item \textbf{Contextual Awareness and Proactive Safety:} A critical feature of the framework is its emphasis on contextual awareness. By understanding the dynamic interactions between the vehicle, its occupants, and the external environment, the system adapts to changing conditions in real time. Proactive safety measures, such as predictive handovers and hazard anticipation, are key to minimizing risks and ensuring user confidence \cite{t68}.
\end{itemize}
\subsubsection{Proposed Generic Algorithm}
As part of the proposed conceptual framework for HVI and AI collaboration, we present a generic algorithm designed for cognitive decision-making during the interpret phase. This algorithm integrates real-time sensor data, human metrics, and historical information to ensure robust contextual awareness and optimal strategy selection \cite{r15}. It begins by normalizing and fusing diverse input streams, including sensor data and human metrics, using probabilistic techniques such as Bayesian networks to create a unified state. Building on this, the algorithm estimates cognitive states—such as human attention, intent, and cognitive load—forming the basis for actionable contextual insights. Predefined decision strategies are then simulated and evaluated against system goals, such as safety and trust, while ethical and bias filters eliminate any strategies that violate fairness or safety constraints. Risk prediction models, including Monte Carlo simulations, further assess the viability of each strategy, complemented by trust assessments that incorporate human feedback to predict trustworthiness \cite{r16}. Finally, the algorithm employs context-aware strategy selection to rank and identify the optimal strategy, balancing risk, trust, and alignment with system objectives.

This comprehensive algorithm, outlined in Algorithm \ref{algo:1}, embodies the principles of adaptive, human-centered decision-making central to HVIs. By addressing critical elements such as ethical considerations, risk evaluation, and trust incorporation, it fosters seamless HAC, ensuring both system performance and responsiveness to human needs in autonomous systems.

This conceptual framework for HVI and AI collaboration provides a structured approach to designing systems that prioritize adaptability, transparency, and safety. By integrating the SIDA cycle with key components that address user trust, ethical considerations, and real-time decision-making, the framework lays the groundwork for effective HAC in automated driving contexts.

\begin{algorithm}
\caption{\textbf{Cognitive Decision-Making in the Interpret Phase}}
\textbf{Require:} $SensorData$, $HumanMetrics$, $SystemGoals$, $HistoricalData$ \\
\textbf{Ensure:} $ContextualAwareness$, $OptimalStrategy$ \\

\begin{algorithmic}[1]
\State \textbf{Step 1: Contextual Awareness Initialization}
    \State Initialize $FusedData = \{\}$
    \State Initialize $ContextualState = \{\}$

\State \textbf{Step 2: Data Fusion}
    \ForAll {data stream in $SensorData$}
        \State Normalize input data stream
        \State Fuse $HumanMetrics$ and $SensorData$ using Bayesian networks to create $UnifiedState$
    \EndFor
    \State Store $FusedData = UnifiedState$

\State \textbf{Step 3: Cognitive State Estimation}
    \State Input $FusedData$ into CognitiveModel (e.g., ACT-R, Neural-Symbolic Systems)
    \State Estimate $HumanAttention$, $DriverIntent$, and $CognitiveLoad$ from $FusedData$
    \State Store $ContextualState = \{Environment, HumanAttention, DriverIntent\}$

\State \textbf{Step 4: Decision Strategy Simulation}
    \State Initialize $CandidateStrategies = [X]$
    \ForAll {strategy in $PredefinedStrategies$}
        \State Simulate strategy outcomes using $ContextualState$ and $HistoricalData$
        \State Evaluate strategy alignment with $SystemGoals$ (e.g., trust, safety)
        \State Add evaluated strategy to $CandidateStrategies$
    \EndFor

\State \textbf{Step 5: Ethical and Bias Filtering}
    \ForAll {strategy in $CandidateStrategies$}
        \If {strategy violates $EthicalConstraints$ (e.g., fairness, safety)}
            \State Remove strategy from $CandidateStrategies$
        \EndIf
    \EndFor

\State \textbf{Step 6: Risk Prediction}
    \ForAll {strategy in $CandidateStrategies$}
        \State Calculate $RiskScore$ using probabilistic models (e.g., Monte Carlo simulation)
    \EndFor

\State \textbf{Step 7: Trust Assessment}
    \ForAll {strategy in $CandidateStrategies$}
        \State Predict $TrustScore$ using $TrustModel$ (e.g., trust prediction models based on human feedback)
    \EndFor

\State \textbf{Step 8: Context-Aware Strategy Selection}
    \State Rank $CandidateStrategies$ based on $RiskScore$, $TrustScore$, and alignment with $SystemGoals$
    \State Select $OptimalStrategy = CandidateStrategies[TopRanked]$

\State \Return $ContextualAwareness$, $OptimalStrategy$
\end{algorithmic}
\label{algo:1}
\end{algorithm}

\section{Conclusion}
\label{sec:5}
This survey has provided a comprehensive overview of HVIs and AI collaboration, focusing on their potential to optimize decision-making in automated driving. By critically analyzing cognitive modeling techniques, identifying challenges, and proposing a novel taxonomy and conceptual framework, this work highlights the complex yet indispensable role of HAC in advancing automated vehicle systems. The review underscores the need for integrated approaches that combine human adaptability with AI precision to address the dynamic and uncertain nature of real-world driving environments.

The study reveals that while significant advancements have been made in cognitive modeling and decision-making frameworks, challenges related to trust, ethical considerations, and adaptability remain unresolved. Bridging these gaps will require innovative methodologies, advanced data fusion techniques, and frameworks that prioritize both human factors and technological efficiency. Moreover, the proposed taxonomy and algorithm structure emphasize the importance of systematic approaches to interpret, simulate, and optimize HAC in automated systems.

\textbf{Concluding Remarks}
\begin{enumerate}
    \item Human-in-the-loop frameworks are vital to ensuring adaptability and trust in automated driving systems, highlighting the irreplaceable role of human cognition in decision-making.
    \item Robust cognitive models that integrate real-time human metrics and sensor data are essential for predicting and responding to dynamic driving scenarios.
    \item Overcoming challenges such as trust calibration, ethical considerations, and unpredictability in human behavior requires interdisciplinary research and innovative solutions.
    \item Emphasis on ethical filtering and bias mitigation is critical for ensuring fairness and public trust in autonomous decision-making systems.
    \item Future research should focus on developing adaptive, context-aware frameworks that enhance human-machine communication and minimize cognitive workload.
    \item Collaboration between academia, industry, and policymakers will be essential to address regulatory, societal, and technical challenges in achieving full automation.
\end{enumerate}

Lastly, the insights and frameworks presented in this review aim to bridge existing gaps in HVI research and guide future advancements in HAC. By addressing the highlighted challenges and leveraging the proposed methodologies, the field can make significant strides toward safer, more efficient, and human-centered automated driving systems.

\section*{Disclosure Statement}
The authors are not aware of any affiliations, memberships, funding, or financial holdings that might be perceived as affecting the objectivity of this review.

\section*{Conflicts of Interest}
The authors declare that they have no conflict of interest.

\bibliographystyle{IEEEtran}
 \bibliography{references}

\begin{thebibliography}{100}
\providecommand{\url}[1]{#1}
\csname url@samestyle\endcsname
\providecommand{\newblock}{\relax}
\providecommand{\bibinfo}[2]{#2}
\providecommand{\BIBentrySTDinterwordspacing}{\spaceskip=0pt\relax}
\providecommand{\BIBentryALTinterwordstretchfactor}{4}
\providecommand{\BIBentryALTinterwordspacing}{\spaceskip=\fontdimen2\font plus
\BIBentryALTinterwordstretchfactor\fontdimen3\font minus \fontdimen4\font\relax}
\providecommand{\BIBforeignlanguage}[2]{{%
\expandafter\ifx\csname l@#1\endcsname\relax
\typeout{** WARNING: IEEEtran.bst: No hyphenation pattern has been}%
\typeout{** loaded for the language `#1'. Using the pattern for}%
\typeout{** the default language instead.}%
\else
\language=\csname l@#1\endcsname
\fi
#2}}
\providecommand{\BIBdecl}{\relax}
\BIBdecl

\bibitem{r3}
Z.~Tan, N.~Dai, Y.~Su, R.~Zhang, Y.~Li, D.~Wu, and S.~Li, ``Human--machine interaction in intelligent and connected vehicles: A review of status quo, issues, and opportunities,'' \emph{IEEE Transactions on Intelligent Transportation Systems}, vol.~23, no.~9, pp. 13\,954--13\,975, 2021.

\bibitem{r5}
R.~Arunthavanathan, Z.~Sajid, F.~Khan, and E.~Pistikopoulos, ``Artificial intelligence--human intelligence conflict and its impact on process system safety,'' \emph{Digital Chemical Engineering}, vol.~11, p. 100151, 2024.

\bibitem{r2}
M.~Ren, N.~Chen, and H.~Qiu, ``Human-machine collaborative decision-making: An evolutionary roadmap based on cognitive intelligence,'' \emph{International Journal of Social Robotics}, vol.~15, no.~7, pp. 1101--1114, 2023.

\bibitem{r7}
S.~Damiani, E.~Deregibus, and L.~Andreone, ``Driver-vehicle interfaces and interaction: where are they going?'' 2009.

\bibitem{t77}
J.~Inga, M.~Ruess, J.~H. Robens, T.~Nelius, S.~Rothfu{\ss}, S.~Kille, P.~Dahlinger, A.~Lindenmann, R.~Thomaschke, G.~Neumann \emph{et~al.}, ``Human-machine symbiosis: A multivariate perspective for physically coupled human-machine systems,'' \emph{International Journal of Human-Computer Studies}, vol. 170, p. 102926, 2023.

\bibitem{t69}
S.~Mahmoud, E.~Billing, H.~Svensson, and S.~Thill, ``Where to from here? on the future development of autonomous vehicles from a cognitive systems perspective,'' \emph{Cognitive Systems Research}, vol.~76, pp. 63--77, 2022.

\bibitem{t70}
D.~Dellermann, A.~Calma, N.~Lipusch, T.~Weber, S.~Weigel, and P.~Ebel, ``The future of human-ai collaboration: a taxonomy of design knowledge for hybrid intelligence systems,'' \emph{arXiv preprint arXiv:2105.03354}, 2021.

\bibitem{t71}
F.~Fernandez, A.~Sanchez, J.~F. Velez, and B.~Moreno, ``Associated reality: A cognitive human--machine layer for autonomous driving,'' \emph{Robotics and Autonomous Systems}, vol. 133, p. 103624, 2020.

\bibitem{t72}
H.~Ning, R.~Yin, A.~Ullah, and F.~Shi, ``A survey on hybrid human-artificial intelligence for autonomous driving,'' \emph{IEEE Transactions on Intelligent Transportation Systems}, vol.~23, no.~7, pp. 6011--6026, 2021.

\bibitem{t73}
K.~Dargahi~Nobari, F.~Albers, K.~Bartsch, J.~Braun, and T.~Bertram, ``Modeling driver-vehicle interaction in automated driving,'' \emph{Forschung im Ingenieurwesen}, vol.~86, no.~1, pp. 65--79, 2022.

\bibitem{r8}
F.~Biondi, I.~Alvarez, and K.-A. Jeong, ``Human--vehicle cooperation in automated driving: A multidisciplinary review and appraisal,'' \emph{International Journal of Human--Computer Interaction}, vol.~35, no.~11, pp. 932--946, 2019.

\bibitem{r123}
Y.~Liu, R.~Zhao, and Y.~Li, ``A preliminary comparison of drivers’ overtaking behavior between partially automated vehicles and conventional vehicles,'' in \emph{Proceedings of the Human Factors and Ergonomics Society Annual Meeting}, vol.~66, no.~1.\hskip 1em plus 0.5em minus 0.4em\relax SAGE Publications Sage CA: Los Angeles, CA, 2022, pp. 913--917.

\bibitem{s20}
J.~Lee, H.~Rheem, J.~D. Lee, J.~F. Szczerba, and O.~Tsimhoni, ``Teaming with your car: Redefining the driver--automation relationship in highly automated vehicles,'' \emph{Journal of cognitive engineering and decision making}, vol.~17, no.~1, pp. 49--74, 2023.

\bibitem{s21}
S.~Wu and J.~Zhang, ``Research on a compound dual innovation capability model of intelligent manufacturing enterprises,'' \emph{Sustainability}, vol.~13, no.~22, p. 12521, 2021.

\bibitem{s22}
D.~Milakis, B.~Van~Arem, and B.~Van~Wee, ``Policy and society related implications of automated driving: A review of literature and directions for future research,'' \emph{Journal of intelligent transportation systems}, vol.~21, no.~4, pp. 324--348, 2017.

\bibitem{s23}
F.~You, H.~Deng, P.~Hansen, and J.~Zhang, ``Research on transparency design based on shared situation awareness in semi-automatic driving,'' \emph{Applied Sciences}, vol.~12, no.~14, p. 7177, 2022.

\bibitem{s97}
G.~Arastoopour~Irgens, N.~C. Chesler, J.~T. Linderoth, and D.~Williamson~Shaffer, ``Data-enabled cognitive modeling: Validating student engineers’ fuzzy design-based decision-making in a virtual design problem,'' \emph{Computer Applications in Engineering Education}, vol.~25, no.~6, pp. 1001--1017, 2017.

\bibitem{s98}
C.~Feher~da Silva and T.~Hare, ``Humans primarily use model-based inference in the two-stage task. nature human behaviour, 4 (10), 1053--1066,'' 2020.

\bibitem{s100}
R.~Rinaldo, B.~Tarigan, and T.~Juliantine, ``Review: The effect of the teaching game for understanding model on cognitive ability. kinestetik: Jurnal ilmiah pendidikan jasmani, 5 (2), 375--380,'' 2021.

\bibitem{f1}
Y.~Forster, S.~Hergeth, F.~Naujoks, M.~Beggiato, J.~F. Krems, and A.~Keinath, ``Learning and development of mental models during interactions with driving automation: A simulator study,'' in \emph{Driving Assessment Conference}, vol.~10, no. 2019.\hskip 1em plus 0.5em minus 0.4em\relax University of Iowa, 2019.

\bibitem{f10}
J.~Lee, H.~Rheem, J.~D. Lee, J.~F. Szczerba, and O.~Tsimhoni, ``Teaming with your car: Redefining the driver--automation relationship in highly automated vehicles,'' \emph{Journal of cognitive engineering and decision making}, vol.~17, no.~1, pp. 49--74, 2023.

\bibitem{f11}
F.~N. Biondi, M.~Lohani, R.~Hopman, S.~Mills, J.~M. Cooper, and D.~L. Strayer, ``80 mph and out-of-the-loop: Effects of real-world semi-automated driving on driver workload and arousal,'' in \emph{Proceedings of the human factors and ergonomics society annual meeting}, vol.~62, no.~1.\hskip 1em plus 0.5em minus 0.4em\relax SAGE Publications Sage CA: Los Angeles, CA, 2018, pp. 1878--1882.

\bibitem{f12}
D.~He and B.~Donmez, ``Influence of driving experience on distraction engagement in automated vehicles,'' \emph{Transportation research record}, vol. 2673, no.~9, pp. 142--151, 2019.

\bibitem{f13}
K.~Lu, J.~Karlsson, A.~S. Dahlman, B.~A. Sj{\"o}qvist, and S.~Candefjord, ``Detecting driver sleepiness using consumer wearable devices in manual and partial automated real-road driving,'' \emph{IEEE transactions on intelligent transportation systems}, vol.~23, no.~5, pp. 4801--4810, 2021.

\bibitem{f14}
Y.~Forster, S.~Hergeth, F.~Naujoks, M.~Beggiato, J.~F. Krems, and A.~Keinath, ``Learning and development of mental models during interactions with driving automation: A simulator study,'' in \emph{Driving Assessment Conference}, vol.~10, no. 2019.\hskip 1em plus 0.5em minus 0.4em\relax University of Iowa, 2019.

\bibitem{f2}
A.~V{\'a}rhelyi, C.~Kaufmann, C.~Johnsson, and S.~Almqvist, ``Driving with and without automation on the motorway--an observational study,'' \emph{Journal of Intelligent Transportation Systems}, vol.~25, no.~6, pp. 587--608, 2021.

\bibitem{f15}
P.~Czech, M.~Braun, U.~Kre{\ss}el, and B.~Yang, ``Behavior-aware pedestrian trajectory prediction in ego-centric camera views with spatio-temporal ego-motion estimation,'' \emph{Machine Learning and Knowledge Extraction}, vol.~5, no.~3, pp. 957--978, 2023.

\bibitem{f16}
W.~Morales-Alvarez, M.~Marouf, H.~H. Tadjine, and C.~Olaverri-Monreal, ``Real-world evaluation of the impact of automated driving system technology on driver gaze behavior, reaction time and trust,'' in \emph{2021 IEEE Intelligent Vehicles Symposium Workshops (IV Workshops)}.\hskip 1em plus 0.5em minus 0.4em\relax IEEE, 2021, pp. 57--64.

\bibitem{f3}
J.~Lee, H.~Rheem, J.~D. Lee, J.~F. Szczerba, and O.~Tsimhoni, ``Teaming with your car: Redefining the driver--automation relationship in highly automated vehicles,'' \emph{Journal of cognitive engineering and decision making}, vol.~17, no.~1, pp. 49--74, 2023.

\bibitem{f17}
A.~Persson, H.~Jonasson, I.~Fredriksson, U.~Wiklund, and C.~Ahlstr{\"o}m, ``Heart rate variability for classification of alert versus sleep deprived drivers in real road driving conditions,'' \emph{IEEE Transactions on Intelligent Transportation Systems}, vol.~22, no.~6, pp. 3316--3325, 2020.

\bibitem{f18}
A.~S. McDonnell, K.~W. Crabtree, J.~M. Cooper, and D.~L. Strayer, ``This is your brain on autopilot 2.0: The influence of practice on driver workload and engagement during on-road, partially automated driving,'' \emph{Human factors}, vol.~66, no.~8, pp. 2025--2040, 2024.

\bibitem{f4}
M.~R. Endsley, ``Autonomous driving systems: A preliminary naturalistic study of the tesla model s,'' \emph{Journal of Cognitive Engineering and Decision Making}, vol.~11, no.~3, pp. 225--238, 2017.

\bibitem{f19}
C.~Peng, N.~Merat, R.~Romano, F.~Hajiseyedjavadi, E.~Paschalidis, C.~Wei, V.~Radhakrishnan, A.~Solernou, D.~Forster, and E.~Boer, ``Drivers’ evaluation of different automated driving styles: Is it both comfortable and natural?'' \emph{Human factors}, p. 00187208221113448, 2024.

\bibitem{f20}
E.~Shi and K.~Bengler, ``Using task switching to explain effects of non-driving related activities on takeover and manual driving behavior following level 3 automated driving,'' in \emph{Adjunct Proceedings of the 14th International Conference on Automotive User Interfaces and Interactive Vehicular Applications}, 2022, pp. 44--47.

\bibitem{f21}
F.~Codevilla, E.~Santana, A.~M. L{\'o}pez, and A.~Gaidon, ``Exploring the limitations of behavior cloning for autonomous driving,'' in \emph{Proceedings of the IEEE/CVF international conference on computer vision}, 2019, pp. 9329--9338.

\bibitem{f22}
W.~Morales-Alvarez, N.~Smirnov, E.~Matthes, and C.~Olaverri-Monreal, ``Vehicle automation field test: Impact on driver behavior and trust,'' in \emph{2020 IEEE 23rd International Conference on Intelligent Transportation Systems (ITSC)}.\hskip 1em plus 0.5em minus 0.4em\relax IEEE, 2020, pp. 1--6.

\bibitem{f23}
F.~Hauer, A.~Pretschner, and B.~Holzm{\"u}ller, ``Re-using concrete test scenarios generally is a bad idea,'' in \emph{2020 IEEE Intelligent Vehicles Symposium (IV)}.\hskip 1em plus 0.5em minus 0.4em\relax IEEE, 2020, pp. 1305--1310.

\bibitem{f24}
A.~V{\'a}rhelyi, C.~Kaufmann, C.~Johnsson, and S.~Almqvist, ``Driving with and without automation on the motorway--an observational study,'' \emph{Journal of Intelligent Transportation Systems}, vol.~25, no.~6, pp. 587--608, 2021.

\bibitem{s107}
M.~Marcano, S.~D{\'\i}az, J.~P{\'e}rez, and E.~Irigoyen, ``A review of shared control for automated vehicles: Theory and applications,'' \emph{IEEE Transactions on Human-Machine Systems}, vol.~50, no.~6, pp. 475--491, 2020.

\bibitem{s108}
M.~Laakasuo, ``Moral uncanny valley revisited--how human expectations of robot morality based on robot appearance moderate the perceived morality of robot decisions in high conflict moral dilemmas,'' \emph{Frontiers in Psychology}, vol.~14, p. 1270371, 2023.

\bibitem{s110}
H.~Gunes, O.~Celiktutan, and E.~Sariyanidi, ``Live human--robot interactive public demonstrations with automatic emotion and personality prediction,'' \emph{Philosophical Transactions of the Royal Society B}, vol. 374, no. 1771, p. 20180026, 2019.

\bibitem{s111}
T.~Miller, ``Explanation in artificial intelligence: Insights from the social sciences,'' \emph{Artificial intelligence}, vol. 267, pp. 1--38, 2019.

\bibitem{s113}
M.~Chen, S.~Nikolaidis, H.~Soh, D.~Hsu, and S.~Srinivasa, ``Trust-aware decision making for human-robot collaboration: Model learning and planning,'' \emph{ACM Transactions on Human-Robot Interaction (THRI)}, vol.~9, no.~2, pp. 1--23, 2020.

\bibitem{s114}
C.~Huang, C.~Lv, P.~Hang, Z.~Hu, and Y.~Xing, ``Human--machine adaptive shared control for safe driving under automation degradation,'' \emph{IEEE Intelligent Transportation Systems Magazine}, vol.~14, no.~2, pp. 53--66, 2021.

\bibitem{s115}
Z.~R. Khavas, S.~R. Ahmadzadeh, and P.~Robinette, ``Modeling trust in human-robot interaction: A survey,'' in \emph{International conference on social robotics}.\hskip 1em plus 0.5em minus 0.4em\relax Springer, 2020, pp. 529--541.

\bibitem{s116}
F.~Rossi and N.~Mattei, ``Building ethically bounded ai,'' in \emph{Proceedings of the AAAI Conference on Artificial Intelligence}, vol.~33, no.~01, 2019, pp. 9785--9789.

\bibitem{s156}
A.~W. Thomas, U.~Lindenberger, W.~Samek, and K.-R. M{\"u}ller, ``Evaluating deep transfer learning for whole-brain cognitive decoding,'' \emph{Journal of the Franklin Institute}, vol. 360, no.~13, pp. 9754--9787, 2023.

\bibitem{s157}
L.~He, H.~Li, S.~K. Holland, W.~Yuan, M.~Altaye, and N.~A. Parikh, ``Early prediction of cognitive deficits in very preterm infants using functional connectome data in an artificial neural network framework,'' \emph{NeuroImage: Clinical}, vol.~18, pp. 290--297, 2018.

\bibitem{s158}
X.~Tong, H.~Xie, N.~Carlisle, G.~A. Fonzo, D.~J. Oathes, J.~Jiang, and Y.~Zhang, ``Transdiagnostic connectome signatures from resting-state fmri predict individual-level intellectual capacity,'' \emph{Translational psychiatry}, vol.~12, no.~1, p. 367, 2022.

\bibitem{s159}
E.~Moradi, M.~Prakash, A.~Hall, A.~Solomon, B.~Strange, J.~Tohka, and A.~D.~N. Initiative, ``Machine learning prediction of future amyloid beta positivity in amyloid-negative individuals,'' \emph{Alzheimer's Research \& Therapy}, vol.~16, no.~1, p.~46, 2024.

\bibitem{s160}
K.~Miller, M.~Botvinick, and C.~Brody, ``From predictive models to cognitive models: Separable behavioral processes underlying reward learning in the rat. biorxiv, 461129,'' 2021.

\bibitem{s161}
S.~Chopra, E.~Dhamala, C.~Lawhead, J.~A. Ricard, E.~R. Orchard, L.~An, P.~Chen, N.~Wulan, P.~Kumar, A.~Rubenstein \emph{et~al.}, ``Reliable and generalizable brain-based predictions of cognitive functioning across common psychiatric illness,'' \emph{medRxiv}, pp. 2022--12, 2022.

\bibitem{s162}
H.~Musto, D.~Stamate, I.~Pu, and D.~Stahl, ``A machine learning approach for predicting deterioration in alzheimer’s disease,'' in \emph{2021 20th IEEE international conference on machine learning and applications (ICMLA)}.\hskip 1em plus 0.5em minus 0.4em\relax IEEE, 2021, pp. 1443--1448.

\bibitem{s163}
K.~C. Bathina, M.~Ten~Thij, L.~Lorenzo-Luaces, L.~A. Rutter, and J.~Bollen, ``Individuals with depression express more distorted thinking on social media,'' \emph{Nature human behaviour}, vol.~5, no.~4, pp. 458--466, 2021.

\bibitem{s164}
S.~Palmqvist, P.~S. Insel, H.~Zetterberg, K.~Blennow, B.~Brix, E.~Stomrud, N.~Mattsson, O.~Hansson, A.~D.~N. Initiative \emph{et~al.}, ``Accurate risk estimation of $\beta$-amyloid positivity to identify prodromal alzheimer's disease: cross-validation study of practical algorithms,'' \emph{Alzheimer's \& Dementia}, vol.~15, no.~2, pp. 194--204, 2019.

\bibitem{s165}
A.~W. Thomas, K.-R. M{\"u}ller, and W.~Samek, ``Deep transfer learning for whole-brain fmri analyses,'' in \emph{OR 2.0 Context-Aware Operating Theaters and Machine Learning in Clinical Neuroimaging: Second International Workshop, OR 2.0 2019, and Second International Workshop, MLCN 2019, Held in Conjunction with MICCAI 2019, Shenzhen, China, October 13 and 17, 2019, Proceedings 2}.\hskip 1em plus 0.5em minus 0.4em\relax Springer, 2019, pp. 59--67.

\bibitem{n25}
R.~M. Van~der Heiden, J.~L. Kenemans, S.~F. Donker, and C.~P. Janssen, ``The effect of cognitive load on auditory susceptibility during automated driving,'' \emph{Human factors}, vol.~64, no.~7, pp. 1195--1209, 2022.

\bibitem{n26}
Y.~Forster, S.~Hergeth, F.~Naujoks, M.~Beggiato, J.~F. Krems, and A.~Keinath, ``Learning and development of mental models during interactions with driving automation: A simulator study,'' in \emph{Driving Assessment Conference}, vol.~10, no. 2019.\hskip 1em plus 0.5em minus 0.4em\relax University of Iowa, 2019.

\bibitem{n27}
N.~Pongsakornsathien, Y.~Lim, A.~Gardi, S.~Hilton, L.~Planke, R.~Sabatini, T.~Kistan, and N.~Ezer, ``Sensor networks for aerospace human-machine systems,'' \emph{Sensors}, vol.~19, no.~16, p. 3465, 2019.

\bibitem{n28}
C.~Lebiere, L.~M. Blaha, C.~K. Fallon, and B.~Jefferson, ``Adaptive cognitive mechanisms to maintain calibrated trust and reliance in automation,'' \emph{Frontiers in Robotics and AI}, vol.~8, p. 652776, 2021.

\bibitem{n29}
R.~K. Steinberg, M.~J. Driggs, and A.~Jackson, ``The economic benefits of human performance models in systems engineering,'' in \emph{INCOSE International Symposium}, vol.~32.\hskip 1em plus 0.5em minus 0.4em\relax Wiley Online Library, 2022, pp. 141--148.

\bibitem{t75}
R.~Meyer, R.~Graf~von Spee, E.~Altendorf, and F.~O. Flemisch, ``Gesture-based vehicle control in partially and highly automated driving for impaired and non-impaired vehicle operators: A pilot study,'' in \emph{Universal Access in Human-Computer Interaction. Methods, Technologies, and Users: 12th International Conference, UAHCI 2018, Held as Part of HCI International 2018, Las Vegas, NV, USA, July 15-20, 2018, Proceedings, Part I 12}.\hskip 1em plus 0.5em minus 0.4em\relax Springer, 2018, pp. 216--227.

\bibitem{n30}
Y.~Li and C.~M. Burns, ``Modeling automation with cognitive work analysis to support human-automation coordination,'' \emph{Journal of Cognitive Engineering and Decision Making}, 2017.

\bibitem{n31}
V.~A. Banks, A.~Eriksson, J.~O'Donoghue, and N.~A. Stanton, ``Is partially automated driving a bad idea? observations from an on-road study,'' \emph{Applied Ergonomics}, 2018.

\bibitem{n32}
L.~Strickland, S.~Farrell, M.~K. Wilson, J.~Hutchinson, and S.~Loft, ``How do humans learn about the reliability of automation?'' \emph{Cognitive Research: Principles and Implications}, 2024.

\bibitem{n33}
S.~Pakhomov, S.~E. Marino, S.~J. Banks, and C.~Bernick, ``Using automatic speech recognition to assess spoken responses to cognitive tests of semantic verbal fluency,'' \emph{Speech Communication}, 2015.

\bibitem{n34}
O.~O. Mwambe, P.~X. Tan, and E.~Kamioka, ``Bioinformatics-based adaptive system towards real-time dynamic e-learning content personalization,'' \emph{Education Sciences}, 2020.

\bibitem{r121}
R.~Madigan, T.~Louw, and N.~Merat, ``The effect of varying levels of vehicle automation on drivers’ lane changing behaviour,'' \emph{PloS one}, vol.~13, no.~2, p. e0192190, 2018.

\bibitem{r122}
M.~Wang, R.~Parikh, and M.~Jeon, ``Using multilevel hidden markov models to understand driver hazard avoidance during the takeover process in conditionally automated vehicles,'' in \emph{Proceedings of the Human Factors and Ergonomics Society Annual Meeting}, vol.~67, no.~1.\hskip 1em plus 0.5em minus 0.4em\relax SAGE Publications Sage CA: Los Angeles, CA, 2023, pp. 698--704.

\bibitem{r124}
A.~A. Kaleefathullah, N.~Merat, Y.~M. Lee, Y.~B. Eisma, R.~Madigan, J.~Garcia, and J.~d. Winter, ``External human--machine interfaces can be misleading: An examination of trust development and misuse in a cave-based pedestrian simulation environment,'' \emph{Human factors}, vol.~64, no.~6, pp. 1070--1085, 2022.

\bibitem{t76}
Y.~Xing, C.~Lv, D.~Cao, and P.~Hang, ``Toward human-vehicle collaboration: Review and perspectives on human-centered collaborative automated driving,'' \emph{Transportation research part C: emerging technologies}, vol. 128, p. 103199, 2021.

\bibitem{r125}
J.~Forke, P.~Fr{\"o}hlich, S.~Suette, M.~Gafert, J.~Puthenkalam, L.~Diamond, M.~Zeilinger, and M.~Tscheligi, ``Understanding the headless rider: Display-based awareness and intent-communication in automated vehicle-pedestrian interaction in mixed traffic,'' \emph{Multimodal Technologies and Interaction}, vol.~5, no.~9, p.~51, 2021.

\bibitem{r126}
M.~Choe, E.~Bosch, J.~Dong, I.~Alvarez, M.~Oehl, C.~Jallais, A.~Alsaid, C.~Nadri, and M.~Jeon, ``Emotion garage vol. iv: Creating empathic in-vehicle interfaces with generative ais for automated vehicle contexts,'' in \emph{Adjunct Proceedings of the 15th International Conference on Automotive User Interfaces and Interactive Vehicular Applications}, 2023, pp. 234--236.

\bibitem{r127}
K.~De~Clercq, A.~Dietrich, J.~P. N{\'u}{\~n}ez~Velasco, J.~De~Winter, and R.~Happee, ``External human-machine interfaces on automated vehicles: Effects on pedestrian crossing decisions,'' \emph{Human factors}, vol.~61, no.~8, pp. 1353--1370, 2019.

\bibitem{r128}
W.~Xue, R.~Zheng, B.~Yang, Z.~Wang, T.~Kaizuka, and K.~Nakano, ``An adaptive model predictive approach for automated vehicle control in fallback procedure based on virtual vehicle scheme,'' \emph{Journal of intelligent and connected vehicles}, vol.~2, no.~2, pp. 67--77, 2019.

\bibitem{r129}
N.~A. Stanton, A.~Eriksson, V.~A. Banks, and P.~A. Hancock, ``Turing in the driver's seat: Can people distinguish between automated and manually driven vehicles?'' \emph{Human Factors and Ergonomics in Manufacturing \& Service Industries}, vol.~30, no.~6, pp. 418--425, 2020.

\bibitem{r130}
P.~Bazilinskyy, S.~M. Petermeijer, V.~Petrovych, D.~Dodou, and J.~C. de~Winter, ``Take-over requests in highly automated driving: A crowdsourcing survey on auditory, vibrotactile, and visual displays,'' \emph{Transportation research part F: traffic psychology and behaviour}, vol.~56, pp. 82--98, 2018.

\bibitem{r131}
Sm, ``How do vehicle automated features help or hurt driving performance?'' \emph{Ergonomics International Journal}, 2023.

\bibitem{r132}
S.~Easa, ``Transportation ergonomics for self-driving automated vehicles: Out-dated or necessary?'' \emph{Ergonomics International Journal}, 2020.

\bibitem{r133}
M.~K{\"o}rber, E.~Baseler, and K.~Bengler, ``Introduction matters: Manipulating trust in automation and reliance in automated driving,'' \emph{Applied ergonomics}, vol.~66, pp. 18--31, 2018.

\bibitem{r134}
N.~Fereydooni, S.~T. Scott-Sharoni, B.~N. Walker, J.~K. Lenneman, B.~P. Austin, and T.~Yoshida, ``The impact of content temporality and modality in automotive user interface on trust and comfort,'' in \emph{Proceedings of the Human Factors and Ergonomics Society Annual Meeting}, vol.~67, no.~1.\hskip 1em plus 0.5em minus 0.4em\relax SAGE Publications Sage CA: Los Angeles, CA, 2023, pp. 1971--1976.

\bibitem{r135}
H.~Guo, L.~Song, J.~Liu, F.-Y. Wang, D.~Cao, H.~Chen, C.~Lv, and P.~C.-K. Luk, ``Hazard-evaluation-oriented moving horizon parallel steering control for driver-automation collaboration during automated driving,'' \emph{IEEE/CAA Journal of Automatica Sinica}, vol.~5, no.~6, pp. 1062--1073, 2018.

\bibitem{r136}
A.~Zgonnikov, D.~Abbink, and G.~Markkula, ``Should i stay or should i go? cognitive modeling of left-turn gap acceptance decisions in human drivers,'' \emph{Human factors}, vol.~66, no.~5, pp. 1399--1413, 2024.

\bibitem{r137}
S.~Hwang and D.~Lee, ``Evaluation of preferred automated driving patterns based on a driving propensity using fuzzy inference system,'' \emph{Journal of Advanced Transportation}, vol. 2024, no.~1, p. 6628559, 2024.

\bibitem{r138}
M.~A. Nees, ``Drivers’ perceptions of functionality implied by terms used to describe automation in vehicles,'' in \emph{Proceedings of the Human Factors and Ergonomics Society Annual Meeting}, vol.~62, no.~1.\hskip 1em plus 0.5em minus 0.4em\relax SAGE Publications Sage CA: Los Angeles, CA, 2018, pp. 1893--1897.

\bibitem{r139}
F.~Walker, A.~Boelhouwer, T.~Alkim, W.~B. Verwey, and M.~H. Martens, ``Changes in trust after driving level 2 automated cars,'' \emph{Journal of Advanced Transportation}, vol. 2018, no.~1, p. 1045186, 2018.

\bibitem{r140}
P.~Joisten, E.~Alexandi, R.~Drews, L.~Klassen, P.~Petersohn, A.~Pick, S.~Schwindt, and B.~Abendroth, ``Displaying vehicle driving mode: Effects on pedestrian behavior and perceived safety,'' in \emph{Human Systems Engineering and Design II: Proceedings of the 2nd International Conference on Human Systems Engineering and Design (IHSED2019): Future Trends and Applications}.\hskip 1em plus 0.5em minus 0.4em\relax Springer, 2020, pp. 250--256.

\bibitem{r142}
S.~Aoki and R.~R. Rajkumar, ``A-drive: Autonomous deadlock detection and recovery at road intersections for connected and automated vehicles,'' in \emph{2022 IEEE Intelligent Vehicles Symposium (IV)}.\hskip 1em plus 0.5em minus 0.4em\relax IEEE, 2022, pp. 29--36.

\bibitem{r143}
J.~Nunez~Velasco, A.~de~Vries, H.~Farah, B.~van Arem, and M.~Hagenzieker, ``Cyclists’ crossing intentions when interacting with automated vehicles: A virtual reality study,'' \emph{Information (Switzerland)}, vol.~12, no.~1, pp. 1--15, 2021.

\bibitem{r144}
T.~Fuest, A.~Feierle, E.~Schmidt, and K.~Bengler, ``Effects of marking automated vehicles on human drivers on highways,'' \emph{Information}, vol.~11, no.~6, p. 286, 2020.

\bibitem{r145}
F.~Hartwich, C.~Hollander, D.~Johannmeyer, and J.~F. Krems, ``Improving passenger experience and trust in automated vehicles through user-adaptive hmis:“the more the better” does not apply to everyone,'' \emph{Frontiers in Human Dynamics}, vol.~3, p. 669030, 2021.

\bibitem{r146}
T.~Sch{\"u}rmann and P.~Beckerle, ``Personalizing human-agent interaction through cognitive models,'' \emph{Frontiers in Psychology}, vol.~11, p. 561510, 2020.

\bibitem{r147}
A.~H. Kalantari, Y.~Yang, Y.~M. Lee, N.~Merat, and G.~Markkula, ``Driver-pedestrian interactions at unsignalized crossings are not in line with the nash equilibrium,'' \emph{IEEE Access}, 2023.

\bibitem{r148}
J.~Hu, B.~T.-W. Lin, J.~H. Vega, and N.~R.-L. Tsiang, ``Predictive models of driver deceleration and acceleration responses to lead vehicle cutting in and out,'' \emph{Transportation research record}, vol. 2677, no.~5, pp. 92--102, 2023.

\bibitem{r149}
J.~F. Fisac, E.~Bronstein, E.~Stefansson, D.~Sadigh, S.~S. Sastry, and A.~D. Dragan, ``Hierarchical game-theoretic planning for autonomous vehicles,'' in \emph{2019 International conference on robotics and automation (ICRA)}.\hskip 1em plus 0.5em minus 0.4em\relax IEEE, 2019, pp. 9590--9596.

\bibitem{r150}
D.~Li, G.~Liu, and B.~Xiao, ``Human-like driving decision at unsignalized intersections based on game theory,'' \emph{Proceedings of the Institution of Mechanical Engineers, Part D: Journal of Automobile Engineering}, vol. 237, no.~1, pp. 159--173, 2023.

\bibitem{r151}
H.~Farah, W.~Daamen, and S.~Hoogendoorn, ``How do drivers negotiate horizontal ramp curves in system interchanges in the netherlands?'' \emph{Safety science}, vol. 119, pp. 58--69, 2019.

\bibitem{r153}
F.~Camara, N.~Bellotto, S.~Cosar, F.~Weber, D.~Nathanael, M.~Althoff, J.~Wu, J.~Ruenz, A.~Dietrich, G.~Markkula \emph{et~al.}, ``Pedestrian models for autonomous driving part ii: high-level models of human behavior,'' \emph{IEEE Transactions on Intelligent Transportation Systems}, vol.~22, no.~9, pp. 5453--5472, 2020.

\bibitem{r154}
I.~G. Jin, B.~Sch{\"u}rmann, R.~M. Murray, and M.~Althoff, ``Risk-aware motion planning for automated vehicle among human-driven cars,'' in \emph{2019 American Control Conference (ACC)}.\hskip 1em plus 0.5em minus 0.4em\relax IEEE, 2019, pp. 3987--3993.

\bibitem{r155}
P.~Kosmides, K.~Demestichas, K.~Avgerinakis, E.~Trouva, S.~Bianchi, A.~Barisone, K.~Risvas, K.~Moustakas, A.~Rodak, M.~Kruszewski \emph{et~al.}, ``Bringing trust to autonomous mobility,'' in \emph{2020 AEIT International Conference of Electrical and Electronic Technologies for Automotive (AEIT AUTOMOTIVE)}.\hskip 1em plus 0.5em minus 0.4em\relax IEEE, 2020, pp. 1--6.

\bibitem{r156}
D.~Rojas~Rueda, M.~J. Nieuwenhuijsen, H.~Khreis, and H.~Frumkin, ``Autonomous vehicles and public health,'' \emph{Annu Rev Public Health. 2020 Apr 2; 41: 329-45}, 2020.

\bibitem{r157}
E.~Portouli, D.~Nathanael, A.~Amditis, Y.~M. Lee, N.~Merat, J.~Uttley, O.~Giles, G.~Markkula, A.~Dietrich, A.~Schieben \emph{et~al.}, ``Methodologies to understand the road user needs when interacting with automated vehicles,'' in \emph{HCI in Mobility, Transport, and Automotive Systems: First International Conference, MobiTAS 2019, Held as Part of the 21st HCI International Conference, HCII 2019, Orlando, FL, USA, July 26-31, 2019, Proceedings 21}.\hskip 1em plus 0.5em minus 0.4em\relax Springer, 2019, pp. 35--45.

\bibitem{r158}
C.~Wu, H.~Wu, N.~Lyu, and M.~Zheng, ``Take-over performance and safety analysis under different scenarios and secondary tasks in conditionally automated driving,'' \emph{IEEE Access}, vol.~7, pp. 136\,924--136\,933, 2019.

\bibitem{r159}
X.~Ji, K.~Yang, X.~Na, C.~Lv, and Y.~Liu, ``Shared steering torque control for lane change assistance: A stochastic game-theoretic approach,'' \emph{IEEE Transactions on Industrial Electronics}, vol.~66, no.~4, pp. 3093--3105, 2018.

\bibitem{r160}
A.~Feierle, S.~Danner, S.~Steininger, and K.~Bengler, ``Information needs and visual attention during urban, highly automated driving—an investigation of potential influencing factors,'' \emph{Information}, vol.~11, no.~2, p.~62, 2020.

\bibitem{r161}
G.~Previati and G.~Mastinu, ``Sumo roundabout simulation with human in the loop,'' in \emph{SUMO Conference Proceedings}, vol.~4, 2023, pp. 29--40.

\bibitem{r162}
A.~Habibovic, V.~M. Lundgren, J.~Andersson, M.~Klingeg{\aa}rd, T.~Lagstr{\"o}m, A.~Sirkka, J.~Fagerl{\"o}nn, C.~Edgren, R.~Fredriksson, S.~Krupenia \emph{et~al.}, ``Communicating intent of automated vehicles to pedestrians,'' \emph{Frontiers in psychology}, vol.~9, p. 1336, 2018.

\bibitem{r167}
S.~Anwer, H.~Li, M.~F. Antwi‐Afari, W.~Umer, and A.~Y.~L. Wong, ``Evaluation of physiological metrics as real-time measurement of physical fatigue in construction workers: State-of-the-art review,'' \emph{Journal of Construction Engineering and Management}, 2021.

\bibitem{r168}
------, ``Cardiorespiratory and thermoregulatory parameters are good surrogates for measuring physical fatigue during a simulated construction task,'' \emph{International Journal of Environmental Research and Public Health}, 2020.

\bibitem{r169}
C.~M.~V. Bustamante, K.~Alama-Maruta, C.~Ng, and D.~D. Coppersmith, ``Should machines be allowed to ‘read our minds’? uses and regulation of biometric techniques that attempt to infer mental states,'' 2022.

\bibitem{r170}
A.~Pimenta, D.~Carneiro, P.~Novais, and J.~Neves, ``Detection of distraction and fatigue in groups through the analysis of interaction patterns with computers,'' in \emph{Intelligent Distributed Computing VIII}.\hskip 1em plus 0.5em minus 0.4em\relax Springer, 2015, pp. 29--39.

\bibitem{t48}
N.~L. Morris, C.~M. Craig, and K.~R. Schwieters, ``Trust in automated vehicles under normal and drowsy driving conditions,'' \emph{Proceedings of the Human Factors and Ergonomics Society Annual Meeting}, 2023.

\bibitem{t49}
W.~Payre, J.~Perelló-March, G.~Sabaliauskaite, H.~Jadidbonab, S.~A. Shaikh, and S.~A. Birrell, ``How system failures and ransomwares affect drivers' trust and attitudes in an automated car? a simulator study,'' in \emph{Proceedings of the AHFE International Conference}, 2022.

\bibitem{t50}
U.~ERGİN, ``One of the first fatalities of a self-driving car: Root cause analysis of the 2016 tesla model s 70d crash,'' \emph{Trafik Ve Ulaşım Araştırmaları Dergisi}, 2022.

\bibitem{t51}
R.~Utriainen and M.~Pöllänen, ``The needed features of connected and automated vehicles to prevent passenger car crashes caused by driving errors,'' \emph{Future Transportation}, 2021.

\bibitem{t52}
W.~Payre, J.~Perellomarch, G.~Sabaliauskaite, H.~Jadidbonab, S.~A. Shaikh, H.~D.~N. Nguyen, and S.~A. Birrell, ``Understanding drivers' trust after software malfunctions and cyber intrusions of digital displays in an automated car,'' in \emph{Proceedings of the AHFE International Conference}, 2022.

\bibitem{r12}
A.~J.~M. Muzahid, S.~F. Kamarulzaman, M.~A. Rahman, S.~A. Murad, M.~A.~S. Kamal, and A.~H. Alenezi, ``Multiple vehicle cooperation and collision avoidance in automated vehicles: Survey and an ai-enabled conceptual framework,'' \emph{Scientific reports}, vol.~13, no.~1, p. 603, 2023.

\bibitem{t53}
J.~v.~d. Sluis, O.~O.~d. Camp, J.~Broos, I.~Yalcinkaya, and E.~d. Gelder, ``Describing i2v communication in scenarios for simulation-based safety assessment of truck platooning,'' \emph{Electronics}, 2021.

\bibitem{t54}
V.~N. Pai, I.~Barosan, and A.~K. Saberi, ``Map and its impact on the functional safety of automated driving vehicles,'' \emph{Journal of Software Engineering for Autonomous Systems}, 2023.

\bibitem{t55}
D.~Hes, R.~Lattarulo, J.~Pérez, T.~Hesse, and F.~Köster, ``Negotiation of cooperative maneuvers for automated vehicles: Experimental results,'' in \emph{Proceedings of the IEEE Intelligent Transportation Systems Conference (ITSC)}, 2019.

\bibitem{t56}
A.~Knauss, C.~Berger, and H.~Eriksson, ``Towards state-of-the-art and future trends in testing of active safety systems,'' in \emph{Proceedings of the ACM/IEEE International Conference on Software Engineering}, 2016.

\bibitem{t57}
J.~Orlovska, F.~Novakazi, C.~Wickman, and R.~Söderberg, ``Mixed-method design for user behavior evaluation of automated driver assistance systems: An automotive industry case,'' \emph{Proceedings of the Design Society International Conference on Engineering Design}, 2019.

\bibitem{t58}
\BIBentryALTinterwordspacing
S.~Li, Y.~Zhang, S.~Edwards, and P.~Blythe, ``Exploration into the needs and requirements of the remote driver when teleoperating the 5g-enabled level 4 automated vehicle in the real world—a case study of 5g connected and automated logistics,'' \emph{Sensors}, vol.~23, no.~2, p. 820, 2023. [Online]. Available: \url{https://doi.org/10.3390/s23020820}
\BIBentrySTDinterwordspacing

\bibitem{t59}
\BIBentryALTinterwordspacing
C.~Gold, M.~Körber, D.~Lechner, and K.~Bengler, ``Taking over control from highly automated vehicles in complex traffic situations,'' \emph{Human Factors: The Journal of the Human Factors and Ergonomics Society}, vol.~58, no.~4, pp. 642--652, 2016. [Online]. Available: \url{https://doi.org/10.1177/0018720816634226}
\BIBentrySTDinterwordspacing

\bibitem{r13}
A.~J.~M. Muzahid, S.~F. Kamarulzaman, and M.~A. Rahman, ``Comparison of ppo and sac algorithms towards decision making strategies for collision avoidance among multiple autonomous vehicles,'' in \emph{2021 International Conference on Software Engineering \& Computer Systems and 4th International Conference on Computational Science and Information Management (ICSECS-ICOCSIM)}.\hskip 1em plus 0.5em minus 0.4em\relax IEEE, 2021, pp. 200--205.

\bibitem{t60}
\BIBentryALTinterwordspacing
M.~Körber, E.~Baseler, and K.~Bengler, ``Introduction matters: manipulating trust in automation and reliance in automated driving,'' \emph{Applied Ergonomics}, vol.~66, pp. 18--31, 2018. [Online]. Available: \url{https://doi.org/10.1016/j.apergo.2017.07.006}
\BIBentrySTDinterwordspacing

\bibitem{t61}
\BIBentryALTinterwordspacing
S.~Sheng, E.~Pakdamanian, K.~Han, Z.~Wang, J.~Lenneman, and F.~Lu, ``Trust-based route planning for automated vehicles,'' \emph{arXiv preprint}, 2021. [Online]. Available: \url{https://doi.org/10.48550/arxiv.2101.03267}
\BIBentrySTDinterwordspacing

\bibitem{r14}
A.~J.~M. Muzahid, S.~F. Kamarulzaman, M.~A. Rahman, and A.~H. Alenezi, ``Deep reinforcement learning-based driving strategy for avoidance of chain collisions and its safety efficiency analysis in autonomous vehicles,'' \emph{IEEE Access}, vol.~10, pp. 43\,303--43\,319, 2022.

\bibitem{t62}
\BIBentryALTinterwordspacing
M.~Hagenzieker, S.~Kint, L.~Vissers, I.~Schagen, J.~Bruin, and P.~G. et~al., ``Interactions between cyclists and automated vehicles: results of a photo experiment,'' \emph{Journal of Transportation Safety \& Security}, vol.~12, no.~1, pp. 94--115, 2019. [Online]. Available: \url{https://doi.org/10.1080/19439962.2019.1591556}
\BIBentrySTDinterwordspacing

\bibitem{t63}
\BIBentryALTinterwordspacing
A.~Feierle, M.~Rettenmaier, F.~Zeitlmeir, and K.~Bengler, ``Multi-vehicle simulation in urban automated driving: technical implementation and added benefit,'' \emph{Information}, vol.~11, no.~5, p. 272, 2020. [Online]. Available: \url{https://doi.org/10.3390/info11050272}
\BIBentrySTDinterwordspacing

\bibitem{t64}
\BIBentryALTinterwordspacing
D.~Hes, R.~Lattarulo, J.~Pérez, T.~Hesse, and F.~Köster, ``Negotiation of cooperative maneuvers for automated vehicles: experimental results,'' in \emph{2019 IEEE Intelligent Transportation Systems Conference (ITSC)}, 2019. [Online]. Available: \url{https://doi.org/10.1109/itsc.2019.8917464}
\BIBentrySTDinterwordspacing

\bibitem{t65}
\BIBentryALTinterwordspacing
J.~Gerdes and S.~Thornton, ``Implementable ethics for autonomous vehicles,'' in \emph{Autonomous Driving}, 2016, pp. 87--102. [Online]. Available: \url{https://doi.org/10.1007/978-3-662-48847-8_5}
\BIBentrySTDinterwordspacing

\bibitem{t66}
\BIBentryALTinterwordspacing
D.~Miller, M.~Johns, B.~Mok, N.~Gowda, D.~Sirkin, and K.~L. et~al., ``Behavioral measurement of trust in automation,'' in \emph{Proceedings of the Human Factors and Ergonomics Society Annual Meeting}, vol.~60, no.~1, 2016, pp. 1849--1853. [Online]. Available: \url{https://doi.org/10.1177/1541931213601422}
\BIBentrySTDinterwordspacing

\bibitem{t68}
\BIBentryALTinterwordspacing
C.~Kettwich, A.~Schrank, and M.~Oehl, ``Teleoperation of highly automated vehicles in public transport: user-centered design of a human-machine interface for remote-operation and its expert usability evaluation,'' \emph{Multimodal Technologies and Interaction}, vol.~5, no.~5, p.~26, 2021. [Online]. Available: \url{https://doi.org/10.3390/mti5050026}
\BIBentrySTDinterwordspacing

\bibitem{r15}
A.~J.~M. Muzahid, M.~A. Rahim, S.~A. Murad, S.~F. Kamarulzaman, and M.~A. Rahman, ``Optimal safety planning and driving decision-making for multiple autonomous vehicles: A learning based approach,'' in \emph{2021 Emerging Technology in Computing, Communication and Electronics (ETCCE)}.\hskip 1em plus 0.5em minus 0.4em\relax IEEE, 2021, pp. 1--6.

\bibitem{r16}
A.~J.~M. Muzahid, S.~F. Kamarulzaman, and M.~A. Rahim, ``Learning-based conceptual framework for threat assessment of multiple vehicle collision in autonomous driving,'' in \emph{2020 Emerging Technology in Computing, Communication and Electronics (ETCCE)}.\hskip 1em plus 0.5em minus 0.4em\relax IEEE, 2020, pp. 1--6.

\end{thebibliography}

\end{document}